\documentclass[journal]{IEEEtran}
\input{decl}

\begin{document}

\title{\huge Pseudo-inverse reconstruction of bandlimited signals
from nonuniform generalized samples with orthogonal kernels}

\author{Nguyen T. Thao,~\IEEEmembership{Member,~IEEE,}
Dominik Rzepka,~\IEEEmembership{Member,~IEEE,} and Marek Mi\'skowicz,~\IEEEmembership{Senior Member,~IEEE} 
\thanks{N. T. Thao is with the Department of Electrical Engineering, The City College of New York, CUNY, New York, USA, email: tnguyen@ccny.cuny.edu.}
\thanks{D. Rzepka and M. Mi\'skowicz are with the Department of Measurement and Electronics, AGH University of Krak\'ow, Krak\'ow, Poland, emails: drzepka@agh.edu.pl, miskow@agh.edu.pl}
\thanks{This research was supported by National Center of Science of Poland
under the grant DEC- 2018/31/B/ST7/03874.}
}
\maketitle

\begin{abstract}
Contrary to the traditional pursuit of research on nonuniform sampling of bandlimited signals, the objective of the present paper is not to find sampling conditions that permit perfect reconstruction, but to perform the best possible signal recovery from any given set of nonuniform samples, whether it is finite as in practice, or infinite to achieve the possibility of unique reconstruction in $L^2(\RR)$. This leads us to consider the pseudo-inverse of the whole sampling map as a linear operator of Hilbert spaces. We propose in this paper an iterative algorithm that systematically performs this pseudo-inversion under the following conditions: (i) the input lies in some closed space $\scA$ (such as a space of bandlimited functions); (ii) the samples are formed by inner product of the input with given kernel functions; (iii) these functions are orthogonal at least in a Hilbert space $\scH$ that contains $\scA$. This situation turns out to appear in certain time encoders that are part of the increasingly important area of event-based sampling. As a result of pseudo-inversion, we systematically achieve perfect reconstruction whenever the samples uniquely characterize the input,  we obtain minimal-norm estimates when the sampling is insufficient, and the reconstruction errors are controlled in the case of noisy sampling. The algorithm consists in alternating two projections according to the general method of projections onto convex sets (POCS) and can be implemented by iterating time-varying discrete-time filtering. We finally show that our signal and sampling assumptions appear in a nontrivial manner in other existing problems of data acquisition. This includes multi-channel time encoding where $\scH$ is of the type $L^2(\RR)^M$, and traditional point sampling with the adoption of a Sobolev space $\scH$. This thus uncovers the unexpected possibility of sampling pseudo-inversion in existing applications, while indicating the potential of our formalism to generate future sampling schemes with systematic pseudo-inverse reconstructions.
\end{abstract}


\section{Introduction}

The reconstruction of bandlimited signals from nonuniform samples is a challenging topic that has been studied since the 50's \cite{Duffin52,Yen56,Benedetto92,Grochenig92b,Feichtinger94,Marvasti01,benedetto2001modern}, although its practical development has remained somewhat limited. But this subject is currently attracting new attention with the increasing trend of event-based sampling in data acquisition \cite{Miskowicz2018,Rzepka18,Alexandru20,Adam21,Florescu2015,GONTIER201463,Rudresh2020}. This approach to sampling has grown in an effort to simplify the complexity of the analog sampling circuits, lower their power consumption and simultaneously increase their precision. This is made possible in particular by the replacement of amplitude encoding by time encoding, which takes advantage of the higher precision of solid-state circuits in time measurement \cite{SZYDUCZYNSKI2023112762}. Time encoding has however induced the use of nonuniform samples that were not commonly studied in the past literature. A well-known example is the time encoder introduced by Lazar and T\'oth in \cite{Lazar04} which acquires the integrals of an input signal over successive nonuniform intervals, based an asynchronous Sigma-Delta modulator (ASDM).

In their most general form, nonuniform samples of an input signal $x$ in some Hilbert space $(\scA,\langle\cdot,\cdot\rangle)$ are scalars of the form
\begin{equation}\label{sample0}
\s_k=\langle x,g_k\rangle,\qquad k\in\Z
\end{equation}
where $\Z$ is some set of indices and $(g_k)\inZ$ is a given family of functions in $\scA$, which we call the sampling kernel functions \cite{eldar2005general}. Reconstructing $x$ from $\vs=(\s_k)\inZ$ can be presented as solving the linear equation $Su=\vs$ where $S$ is the linear operator
\begin{equation}\label{map}
S:~u\in\scA~\mapsto~\big(\langle u,g_k\rangle\big)\inZ.
\end{equation}
When $\scA$ is a set of bandlimited functions and $\Z$ is finite, a basic approach is to reduce $S$ to a matrix and estimate $x$ by $S^+\vs$, where $S^+$ is the matrix pseudo-inverse of $S$ \cite{wei2004signal}. This problem reduction to finite-dimensional linear algebra is however incompatible with the theoretically infinite time support of bandlimited signals. Even in practice where signals are always time limited, their time support are typically seen in signal processing as virtually infinite compared to the windows of operation. It was proposed to split the resolution of $Su=\vs$ by finite blocks of signals, in which exact algebraic inversions are performed \cite{Lazar08}. With bandlimited signals, block truncations however creates analytically uncontrolled distortions at the block boundaries and necessitates ad-hoc empirical methods of compensations. This also departs from traditional signal processing which on the contrary preserves input signals in their entirety, while performing finite-complexity approximations of the ideal operations by sliding-window processing (such as FIR filtering). In traditional LTI processing, errors of approximation are well controlled by Fourier analysis. In the context of nonuniform sampling, Fourier analysis is no longer applicable by loss of time invariance. However, linear operations are still rigorously analyzed by functional analysis. With such an approach, the first rigorous numerical method of bandlimited interpolation of nonuniform point samples, called the frame algorithm, was derived by Duffin and Schaeffer \cite{Duffin52}. The method used by Lazar and T\`oth in \cite{Lazar04} for input reconstruction from integrals has a similar structure, based on an algorithm by Feichtinger and Gr\"ochenig \cite{Feichtinger94}. It was shown in \cite{Thao21a,Thao22a} all these types of algorithm can be be converted into iterative time-varying linear filtering, with potential of sliding-window implementations.

Until recently, these iterative methods have been limited to cases where $S$ is exactly invertible, based on some sufficient (and not always necessary) sampling conditions. The goal of this paper is to find iterative algorithms of similar type that converge more generally to the pseudo-inverse $S^\dagger$ of $S$ in the theoretical sense of linear operators in Hilbert spaces. The motivation is to achieve perfect reconstruction whenever $S$ is effectively invertible, while proposing an optimal reconstruction whenever $S$ is not invertible (in case of insufficient sampling) or the sampling sequence $(\s_k)\inZ$ is corrupted by noise. In this way, a goal is to incorporate within a single method an algorithm whose behavior is consistent with the theoretical results of sampling from harmonic analysis while giving optimal solutions in real practical situations of sampling.

Solving this question in the most general case of nonorthogonal sampling kernels $(g_k)\inZ$ is difficult. At the other extreme, reconstructing $x$ from the samples of \eqref{sample0} is trivial when the family $(g_k)\inZ$ is an orthogonal basis of the input space $\scA$. This is the case of Shannon's sampling theorem. In this paper, we consider an intermediate situation where achieving the pseudo-inverse $S^\dagger$ becomes possible by successive filtering, and which is described by the following condition:
\begin{equation}\label{cond0}
\begin{array}{c}
\mbox{$(g_k)\inZ$ is orthogonal in a Hilbert space $(\scH,\langle\cdot,\cdot\rangle)$}\\[0.5ex]
\mbox{and $x$ is in a closed subspace $\scA\subset\scH$.}
\end{array}
\end{equation}
Although seemingly ideal, this condition turns out to be realized in the time-encoding system of Lazar and T\`oth \cite{Lazar04}, in the case where $\scH=L^2(\RR)$ and $\scA$ is a subspace of bandlimited signals. This was noticed and utilized in \cite{Thao21a} to construct an algorithm achieving $S^\dagger$ in the specific application of \cite{Lazar04}. The algorithm was based on a particular application of the method of projection onto convex sets (POCS) \cite{Combettes93,Bauschke96}. This was later generalized to integrate-and-fire encoding with leakage in \cite{Thao22a}. The purpose of the present article is to extract from \cite{Thao21a,Thao22a} the most general framework of pseudo-inversion of $S$ by successive filtering under the abstract assumption of \eqref{cond0}. In this generalization, the content of these references is revisited and reformulated to reach its most fundamental ingredients and obtain a self-sufficient theory that is independent of the applications. Our formalism contains theoretical results as well as efficient techniques of practical implementations. A goal is to propose a new framework of nonuniform sampling schemes for which a pseudo-inverse input reconstruction method by successive filtering is readily available. While this objective is meant to influence the design of future sampling schemes, we also show in this paper an immediate impact of the proposed theory by applying it to two existing sampling/reconstruction schemes: one in multi-channel time encoding \cite{Adam20b,Adam21} and one in the original case of nonuniform point sampling \cite{Grochenig92b}. In these two sampling applications, the authors studied their proposed reconstruction algorithms under specific assumptions of unique reconstruction. In both cases, we show that their own algorithms coincide with our generic POCS algorithm. In this process, we end up pointing previously unknown properties that their algorithm possess, including their ability to achieve perfect reconstruction even in situations where proofs of unique reconstruction are not available, the characteristics of their limit when the sampling is insufficient, and their behavior towards sampling noise. But on the theoretical side, an important role of these two applications is to show that the abstract condition \eqref{cond0} can be found in unexpected situations, using some non-standard technique of signal analysis. In the first example, condition \eqref{cond0} is extracted after some non-trivial reduction of the complex algorithm of \cite{Adam20b,Adam21}. Beyond pointing out the unknown properties of their method, our high-level formalization allows a concise reformulation of it together with an organized presentation of its implementation at the level of discrete-time filters. For a complementary demonstration, the difficulty of the second example is not in the complexity of the sampling system, but in the non-trivial signal theoretic approach that is required. For condition \eqref{cond0} to be realized in this case, the traditional Hilbert space $L^2(\RR)$ needs to be replaced by the homogeneous Sobolev space $\sob$ \cite{grafakos2014modern,devore1993constructive}. This only allows us to prove convergence up to a constant component but does lead for the first time to a result of pseudo-inversion of point sampling by successive filtering.

The paper is organized as follows. We start in Section \ref{sec:gen-samp} by reviewing the basic knowledge that samples of the form \eqref{sample0} bring about an input signal $x$ in a general Hilbert space $\scH$ and without any assumption on the sampling kernels $(g_k)\inZ$. We give the basic principle of the POCS algorithm for finding estimates that are consistent with the samples of \eqref{sample0}. In Section \ref{sec:ortho-POCS}, we show how  condition \eqref{cond0} leads a specific configuration of POCS algorithm that is more efficient and that will be later shown to have special connections with the pseudo-inversion of $S$. For that purpose, we devote Section \ref{sec:pseudo} to reviewing the notion of pseudo-inverse for linear operators in infinite dimension, which is not commonly used knowledge in signal processing. Section \ref{sec:POCSlim} then contains the major mathematical contribution of this article. By starting from a zero initial estimate, we prove that the POCS iteration tends to $S^\dagger\vs$ by {\em contraction} whenever the sampling configuration theoretically allows a stable consistent reconstruction (which is systematically the case when $\Z$ is finite). When the initial estimate is a signal $u\up{0}\neq0$, we show that the POCS limit is more generally the signal of the type $S^\dagger\vs+v$ that is closest to $u\up{0}$ under the constraint $v\in\scA$ and $Sv=0$. This is of particular interest when consistent reconstruction is not unique due to insufficient sampling, and one wishes to pick a consistent estimate that is close to a signal guess of statistical or heuristic nature \cite{Rzepka18}. This reconstruction simultaneously takes care of sampling errors with an action of ``noise shaping'' in the case of oversampling \cite{norsworthy1997delta}. In Section \ref{sec:bin}, we discuss some important aspects of practical implementations. We finally present in Sections \ref{sec:multi-TEM} and \ref{sec:Groch} the two mentioned examples of application.

\section{Consistent reconstruction from generalized samples}\label{sec:gen-samp}

In this section, we review the background on the use of POCS for the reconstruction of bandlimited signals from non-uniform generalized samples. Without loss of generality, we assume that the considered bandlimited signals have Nyquist period 1 and call their space $\scB$.

\subsection{Nonuniform generalized samples}

To understand the specific contribution of POCS to signal reconstruction from samples, we first need to give the most general definition of signal sampling. Let $x$ be a function in some Hilbert space $\scH$ equipped with an inner product $\langle\cdot,\cdot\rangle$. We call a generalized sample of $x\in\scH$ any scalar value of the type
\begin{equation}\nonumber
\s=\langle x,g\rangle
\end{equation}
where $g$ is a known function of $\scH$ \cite{Dvorkind09}, which we call a sampling kernel function. In the most basic example of bandlimited signal sampling, the Hilbert space is $\scH=\scB$ equipped with the canonical inner product $\langle\cdot,\cdot\rangle_2$ of $L^2(\RR)$. The point sample of $x\in\scB$ at instant $\tau$ yields the form
\begin{equation}\label{point-sample}
x(\tau)=\langle x,g\rangle\qquad\mbox{with}\qquad g(t):=\sinc(t-\tau)
\end{equation}
where
\begin{equation}\label{sinc}
\sinc(t):=\sin(\pi t)/(\pi t).
\end{equation}
The point sample of the $n$th derivative of $x\in\scB$ at $\tau$ is also viewed as a generalized sample of $x$ as it can be shown using integration by parts that
$$\midfrac{\dif^n x}{\dif t^n}(\tau)=\langle x,g\rangle\qquad\mbox{with}\qquad g(t):=(-1)^n\midfrac{\dif^n \sinc}{\dif t^n}(t-\tau).$$
We also trivially have generalized samples by integration as illustrated by the following example,
$$\textstyle\int_a^b x(t)\,\dif t=\langle x,g\rangle\qquad\mbox{with}\qquad g(t):=1_{[a,b]}(t)$$
where $1_{[a,b]}(t)$ is the indicator function of $[a,b]$.

When $\scH$ is a space of functions on $\RR$, we say that a set of samples
$(\s_k)\inZ=(\langle x,g_k\rangle)\inZ$ is uniform when $\Z=\ZZ$ and  there exists a function $g\in\scH$ and a period $T$ such that $g_k(t)=g(t-kT)$ for all $k\in\ZZ$. Shannon's sampling theorem applies to the particular case where $g(t)=\sinc(t)$ and $T=1$ in our bandlimited setting. Nonuniform sampling then consists in all of the other cases. This could be when $g_k(t)=g(t-t_k)$ where the instants $t_k$ are not regularly spaced, or when $(g_k)\inZ$ are just not the shifted versions of a single function.

\subsection{Set theoretic view of sampling}

The output of each sample $\s_k=\langle x,g_k\rangle$ gives us the deterministic knowledge that $x$ belongs to the hyperplane
\begin{equation}\label{Hk}
\scH_k:=\big\{v\in\scH:\langle v,g_k\rangle=\s_k\big\}.
\end{equation}
Thus, a full sampling sequence $\vs=(\s_k)\inZ$ tells us that $x$ belongs to the intersection
\begin{equation}\label{Hs}
{\scH_\vs}:=\textstyle\bigcap\limits\inZ\scH_k.
\end{equation}
Perfect reconstruction is possible if and only if ${\scH_\vs}$ is limited to a singleton. When this is not the case, there is no deterministic knowledge from $(\s_k)\inZ$ to distinguish $x$ from any other element of ${\scH_\vs}$. We call the functions of ${\scH_\vs}$ the {\em consistent estimates} of $x$. The next proposition gives some general knowledge on the algebraic structure of $\scH_\vs$.
\line
\begin{proposition}\label{prop:consistent}
Let $\vs=(\s_k)\inZ$ be any sequence. If $\scH_\vs\neq\emptyset$, then, for any $u\in\scH_\vs$,
\begin{equation}\label{Hs-struct}
\scH_\vs=u+\scG^\perp
\end{equation}
where $\scG$ is the closed linear span of $(g_k)\inZ$
$$\scG:=\overline{\mathrm{span}}(g_k)\inZ$$
and $\scG^\perp$ is its orthogonal complement in $\scH$.
\end{proposition}
\line
\begin{IEEEproof}
By assumption, $\s_k=\langle u,g_k\rangle$ for each $k\in\Z$. Hence, $v\in\scH_k$ $\Leftrightarrow$ $\langle v,g_k\rangle=\s_k=\langle u,g_k\rangle$ $\Leftrightarrow$ $\langle v{-}u,g_k\rangle=0$. Thus, $v\in\scH_\vs$ if and only if $v{-}u\in\scG^\perp$.
In other words, $\scH_\vs= u+\scG^\perp$.
\end{IEEEproof}
\ppnoi
Since a linear subspace always contains the 0 vector, then $\scH_\vs$ is a singleton if and only if $\scG^\perp=\{0\}$. This is equivalent to $\scG=\scH$, which means that the linear span of $(g_k)\inZ$ is dense in the whole space $\scH$. Thus, the unique reconstruct of $x$ solely depends on the sampling kernels $(g_k)\inZ$, and not on the input $x$ itself.

\subsection{Systematic estimation}

As mentioned in the introduction, the objective of this paper is not to study the question of unique reconstruction of an input $x$ from its samples. The goal is to perform the best possible approximation of $x$ from given samples $\vs$, whatever they are. While the term of ``best possible'' would require some definition, it is at least intuitive that any reconstruction of $x$ that is not consistent with its samples cannot be optimal. This idea is in fact rationally supported by the following property.
\line
\begin{proposition}\label{prop:convproj}
Let $\scC$ be a closed affine subspace of $\scH$ that contains $x$. Then,
\begin{equation}\label{eq:convproj}
\forall u\in\scH,\qquad\|P_\scC u-x\|\leq\|u-x\|
\end{equation}
where $P_\scC$ designates the orthogonal projection of $\scH$ onto $\scC$ and $\|\cdot\|$ is the norm induced by the inner product $\langle\cdot,\cdot\rangle$ of $\scH$. Moreover, the inequality is {\em strict} whenever $u\notin\scC$.
\end{proposition}
\line
\begin{IEEEproof}
For $u\in\scH$ given, $P_\scC u$ is the unique element $v\in\scC$ such that $u{-}v \perp v{-}w$ for all $w\in\scC$. By the Pythagorean theorem, $\|u{-}w\|^2=\|u{-}v\|^2+\|v{-}w\|^2\geq\|v{-}w\|^2$ with a strict inequality when $u\neq v$, which happens whenever $u\notin\scC$. The result of \eqref{eq:convproj} is the particular case $w=x$.
\end{IEEEproof}
\ppnoi
Then, a systematic reconstruction procedure is to pick the consistent estimate
\begin{equation}\label{sys-rec}
\hat x:=P_{\scH_\vs}u\up{0}
\end{equation}
where $u\up{0}$ is some initial estimate proposed by the user. While $u\up{0}$ can be obtained by heuristic or statistical means, $\hat x$ is an estimate of $x$ that is guaranteed to be better than $u\up{0}$ and cannot be further improved deterministically. It is also the consistent estimate that is closest to $u\up{0}$ with respect to $\|\cdot\|$. The strength of this procedure is that whenever uniqueness of reconstruction is effective, whether one is able to prove it or not, $\hat x$ is guaranteed to be the perfect reconstruction of $x$. In the case of non-unique reconstruction, the type \eqref{sys-rec} of reconstruction was first considered by Yen in \cite{Yen56} for the estimation of a bandlimited signal of $L^2(\RR)$ from a finite number of point samples, with the specific choice of $u\up{0}=0$. In this case, $\hat x$ is the consistent estimate that is closest to 0, and hence of minimum norm. It can be easily shown from the knowledge of \eqref{Hs-struct} that this $\hat x$ must be in $\scG$. We note here that \cite{eldar2005general} studied the case where consistent reconstruction is constrained to a linear subspace that may be different from $\scG$.

The next proposition gives a more analytical description of $P_{\scH_\vs}$.
\line
\begin{proposition}\label{prop:proj}
If $x\in\scH_\vs$, then
\begin{equation}\label{eq:proj}
\forall u\in\scH,\qquad P_{\scH_\vs}u=u+P_\scG(x-u)
\end{equation}
where $\scG$ was defined in Proposition \ref{prop:consistent}.
\end{proposition}
\line
\begin{IEEEproof}
Let $v:=u-P_\scG(x{-}u)$. While $v\in u+ \scG$,  $v=x+(u{-}x)-P_\scG(u{-}x)\in x+\scG^\perp$. So $v$ is the orthogonal projection of $u$ onto $x+\scG^\perp=\scH_\vs$.
\end{IEEEproof}
\ppnoi
Note that while $x$ is unknown, $P_\scG(x{-}u)$ can still be theoretically evaluated as it can be shown to only depend on $u$ and $(\s_k)\inZ$.

\subsection{POCS algorithm}\label{sub:POCSalg}

In the most general case, there is unfortunately  no closed form expression for $P_\scG u$. When $\Z$ is a small finite set, $P_\scG$ can only be obtained by inversion (or pseudo-inversion) of a matrix whose ill conditioning in nonuniform sampling rapidly grows with the size of $\Z$ \cite{Choi98}. The basic technique used in this paper to find $P_{\scH_\vs}u$ is the POCS method \cite{Combettes93,Bauschke96} whose basic function is to retrieve an element in the intersection of $N$ closed convex sets.  As closed affine spaces, the sets $\scH_k$ of \eqref{Hk} are a particular case of closed convex sets. Assuming a finite set $\Z=\{1,{\cdots},N\}$, the basic version of the POCS method applied to \eqref{Hs} consists in the iteration
 \begin{equation}\label{POCS}
u\up{n+1}:=P_{\scH_N}\cdots P_{\scH_1}u\up{n},\qquad n\geq0.
\end{equation}
Since $x$ belongs to every set $\scH_k$, we conclude from Proposition \ref{prop:convproj} that the estimate error $\|u\up{n}\!-x\|$ strictly decreases with $n$ as long as $u\up{n}$ has not reached ${\scH_\vs}$. It is actually known \cite[\S III.B]{Combettes93} that not only $u\up{n}$ always eventually converges to an element $u\up{\infty}$ in ${\scH_\vs}$ in norm, but the limit is more precisely
$$u\up{\infty}=P_{\scH_\vs} u\up{0}.$$
As an intersection of closed affine subspaces, note that ${\scH_\vs}$ is itself a closed affine subspace. Thus, $u\up{\infty}$ is the consistent estimate of $x$ that is closest to the initial iterate $u\up{0}$.

\subsection{Kaczmarz algorithm}\label{sub:Kaczmarz}

The Kaczmarz algorithm is the specific name that is given to the iteration of \eqref{POCS} when the sets $\scH_k$ are simply hyperplanes, as is the case of \eqref{Hk}. Given the explicit description of \eqref{Hk}, $P_{\scH_k}$ yields the following simple expression \cite[p.403]{Bauschke96}
\begin{equation}\label{PSk}
\forall u\in\scH,\qquad P_{\scH_k}u= u+\midfrac{\s_k-\langle u,g_k\rangle}{\| g_k\|^2}\, g_k
\end{equation}
(this can also be obtained from \eqref{eq:proj} in the case where $\scG$ is the linear span of $g_k$ alone). The Kaczmarz algorithm was first used in sampling by Yeh and Stark in \cite{Yeh90} for solving numerically the problem of Yen in \cite{Yen56}. Since then however, this method has not known much development in nonuniform sampling of bandlimited signals, primarily due to its slow convergence.

A randomized version of the Kaczmarz algorithm was later introduced in \cite{strohmer2009randomized} for potential statistical accelerations of the convergence. A standard version of it consists in performing a random permutation of the $N$ projections of \eqref{POCS} at each iteration. But this variant is till not suitable for real-time causal signal processing.

\section{Orthogonal sampling kernels}\label{sec:ortho-POCS}

The slow convergence of the Kaczmarz method is primarily due to the non-orthogonality of the sampling kernels $(g_k)\inZ$. Another fundamental shortcoming of this method is the impossibility to reduce it to an iteration of the type $u\up{n+1}=R u\up{n}$ with some fixed transformation $R$ when $\Z$ is infinite. While the number of samples is always finite in practice, an infinite index set $\Z$ is always needed when dealing with the theoretical question of perfect reconstruction of a bandlimited signal in $L^2(\RR)$. Under the assumption of condition \eqref{cond0} where $\scA$ is the input space, we show in this section that there is a way to reduce the POCS algorithm to alternating two projections only. This even allows $\Z$ to be infinite.
After giving some practical examples of this situation in data acquisition, we present this special POCS algorithm and its properties in absence of noise.

\subsection{Practical examples of orthogonal sampling kernels}

The cases of orthogonal kernels that have appeared in the literature in nonuniform sampling until now \cite{Thao21a,Thao22a} take place in the space $\scH:=L^2(\RR)$ with $\scA:=\scB$ as the closed subspace of inputs. Consider a sampling scheme where the samples of $x\in\scB$ are of the type
\begin{equation}\label{gen-IFS}
\s_k:=\int_{t_{k-1}}^{t_k}f_k(t)\,x(t)\,\dif t,\qquad k\in\Z
\end{equation}
for some increasing sequence of time instants $(t_k)\inZ$ and some family of functions $(f_k)\inZ$ in $L^2(\RR)$. This takes the form of \eqref{sample0} with
$$g_k(t):=f_k(t)\,1_{[t_{k-1},t_k]}(t),\qquad k\in\Z.$$
Clearly, the functions $(g_k)\inZ$ are orthogonal since their supports do not overlap (up to discrete points). The samples of \eqref{gen-IFS} are those of  leaky integrate-and-fire encoding (LIF) when the functions $(f_k)\inZ$ are of the type
$$f_k(t):=e^{-\alpha(t-t_{k-1})}$$
for some constant $\alpha>0$ \cite{Thao22a}. In the case $\alpha=0$, the samples are of the simple form
\begin{equation}\label{ASDM}
\s_k:=\int_{t_{k-1}}^{t_k}x(t)\,\dif t,\qquad k\in\Z
\end{equation}
which are also the type of samples that one extracts from integrate-and-fire \cite{Miskowicz2018,Alexandru20,Adam21,Florescu2015,GONTIER201463,Adam20} or from an asynchronous Sigma-Delta modulator (ASDM) \cite{Lazar04,Thao21a}.

\subsection{Consistent reconstruction}

Under the new assumptions, $x$ belongs to both $\scA$ and $\scH_\vs$ when $\vs=(\s_k)\inZ$ is obtained from \eqref{sample0}. So the set of consistent estimates is
\begin{align}\label{sol-set4}
\scA_\vs:=\scA\cap\scH_\vs.
\end{align}
We are going to see that $\scA_\vs$ has a similar structure to $\scH_\vs$ within $\scA$. Let
\begin{equation}\label{tgk}
 {\widetilde g}_k:=P_\scA g_k,\qquad k\in\Z.
\end{equation}
Since  $g_k-\widetilde g_k$ is orthogonal to $\scA$, then $\langle u,g_k-\widetilde g_k\rangle=0$ for any $u\in\scA$. Thus,
\begin{equation}\label{inner-proj}
\forall u\in\scA,~\forall v\in\scH,\qquad\langle u,g_k\rangle=\langle u,\widetilde g_k\rangle.
\end{equation}
It then follows from \eqref{Hs} and \eqref{Hk} that
\begin{align}
\hspace{-4mm}{\scA_\vs}:=\textstyle\bigcap\limits\inZ\scA_k\quad\mbox{where}\quad\scA_k:\hspace{-0.9mm}
&=\scA\cap\scH_k\label{As}\\[-2ex]
&=\big\{v\in\scA:\langle v,\widetilde g_k\rangle=\s_k\big\}.\nonumber
\end{align}
With a proof similar to that of Proposition \ref{prop:consistent}, we have the following result.
\line
\begin{proposition}\label{prop:consistent2}
Let $\vs=(\s_k)\inZ$ be any sequence. If $\scA_\vs\neq\emptyset$, then, for any $u\in\scA_\vs$,
\begin{equation}\label{As-set}
\scA_\vs=u+\scF^\perp
\end{equation}
where $\scF$ is the closed linear span of $(\widetilde g_k)\inZ$
\begin{equation}\label{F}
\scF:=\overline{\mathrm{span}}(\widetilde g_k)\inZ
\end{equation}
and $\scF^\perp$ is the orthogonal complement of $\scF$ in $\scA$.
\end{proposition}
\ppnoi
Again, the reconstruction of $x$ is unique if and only if $\scF^\perp=\{0\}$. As this is equivalent to $\scF=\scA$, this means that the linear span of $(\widetilde g_k)\inZ$ is dense in the whole space $\scA$.

\subsection{POCS for orthogonal sampling kernels}\label{subsec:ortho-POCS}

The main point of the previous section was to prepare the detailed notation of the new sampling framework and reformulate the theoretical condition of unique reconstruction. Now, the new contribution is that the POCS method can be directly applied to the 2-set decomposition of \eqref{sol-set4}. This is because $P_{\scH_\vs}$ is now accessible via its expression of \eqref{eq:proj}. Indeed, since $(g_k)\inZ$ is now an {\em orthogonal basis} of $\scG$, we have the explicit expansion of $P_\scG$,
\begin{equation}\label{PL}
\hspace{-2mm}\forall u\in\scH,\quad P_\scG u=\smallsum{k\in\Z}\Big\langle u,\midfrac{ g_k}{\|g_k\|}\Big\rangle\,\midfrac{ g_k}{\|g_k\|}
=\smallsum{k\in\Z}\midfrac{\langle u,g_k\rangle}{\| g_k\|^2}\, g_k.
\end{equation}
It then results from \eqref{eq:proj} and \eqref{sample0} that
\begin{equation}\label{PGs}
\forall u\in\scH,\qquad P_{\scH_\vs}u=u+\smallsum{k\in\Z}\midfrac{\s_k-\langle u,g_k\rangle}{\| g_k\|^2}\, g_k.
\end{equation}
We can thus implement the POCS iteration
\begin{equation}\label{x-iter1}
u\up{n+1}=P_\scA P_{\scH_\vs}u\up{n},\qquad n\geq0.
\end{equation}
As a consequence of Section \ref{sub:POCSalg}, we know that $\|u\up{n}\!-x\|$ strictly decreases as long as $u\up{n}\notin\scA_\vs$, and
\begin{equation}\label{x-inf2}
u\up{\infty}=P_{\scA_\vs}\,u\up{0}.
\end{equation}
But the new contribution is double: while the POCS iteration of \eqref{x-iter1} is reduced to two projections only, it simultaneously allows $\Z$ to be infinite! This is crucial for achieving perfect reconstruction in infinite-dimensional spaces such as $L^2(\RR)$.

\subsection{Relaxed projections}\label{subsec:relax}

\begin{figure}
\vspace{-5mm}
\centerline{\hbox{\scalebox{1.2}{\includegraphics{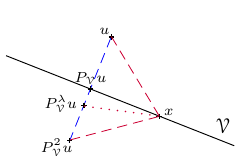}}}}
\caption{Illustration of $P_\scC u$ and $P_\scC^\lambda u$ for $\lambda\in(1,2)$.}\label{fig:proj}
\end{figure}

The estimate error reduction of Proposition \ref{prop:convproj} has in fact the following generalized version.
\line
\begin{proposition}\cite{Bauschke96}\label{prop:relax}
Let $\scC$ be a closed affine subspace of $\scH$ that contains $x$ and
\begin{equation}\label{relaxed-P}
P_\scC^\lambda u:=u+\lambda(P_\scC u-u)
\end{equation}
for any $\lambda\in\RR$ and $u\in\scH$. Then,
\begin{equation}\label{eq:relax}
\forall\lambda\in[0,2],~\forall u\in\scH,\qquad\|P^\lambda_\scC u-x\|\leq\|u-x\|.
\end{equation}
Moreover, the inequality is {\em strict} when $\lambda\in(0,2)$ and $u\notin\scC$.
\end{proposition}
\ppnoi
We illustrate this result graphically in Fig. \ref{fig:proj}.
The parameter $\lambda$ is called a {\em relaxation coefficient.} By iterating
\begin{equation}\label{x-iter3}
u\up{n+1}=P_\scA P^{\lambda_n}_{\scH_\vs}u\up{n},\qquad n\geq0
\end{equation}
where $(\lambda_n)_{n\geq0}$ is some sequence of coefficients in $(0,2)$, one draws the same conclusion as in Section \ref{subsec:ortho-POCS}: assuming that $x\in\scA_\vs$, the estimate error $\|u\up{n}-x\|$ strictly decreases with $n$ as long as $u\up{n}$ has not reached ${\scA_\vs}$. To the best of the authors knowledge, the actual convergence in norm of such a sequence $u\up{n}$ to a point of $\scS$ is not explicitly discussed in the literature. But this can be at least deduced from results in \cite{bauschke2017convex}, which leads to the following theorem.
\line
\begin{theorem}\label{theo:relax}
Assume that  $\scA_\vs\neq\emptyset$ and $(\lambda_n)_{n\geq0}$ is a sequence of coefficients in $[0,2]$ such that $\sum_{n\geq0}\lambda_n(2{-}\lambda_n)=+\infty$. Then the relaxed POCS iteration of \eqref{x-iter3} starting with $u\up{0}\in\scA$ yields the limit of \eqref{x-inf2}.
\end{theorem}
\ppnoi
We justify this in Appendix \ref{app:relax}. The additional relaxation freedom allows in practice to accelerate the convergence. There is no analytical result on the optimal relaxation coefficients, which in practice are typically found empirically.

\section{Sampling operator and pseudo-inverse}\label{sec:pseudo}

We have assumed until now that the sampling sequence $\vs=(\s_k)\inZ$ is given exactly by \eqref{sample0}. In practice however, samples are often corrupted by noise. The question is what happens to the POCS iteration of \eqref{x-iter1} when $\vs$ is deviated by some error. The worst case is when $\vs$ can no longer be the sampled version of any signal in $\scA$. We will see that the POCS iteration is guaranteed to converge with any $\vs$ (up to some theoretical condition on its norm) when the linear operator $S$ of \eqref{map} that maps any $x\in\scA$ into the sequence $(\langle x,g_k\rangle)\inZ$ has a pseudo-inverse $S^\dagger$. In this case, the limit $u\up{\infty}$ will appear to be exactly $S^\dagger\vs$ when $u\up{0}=0$. For the moment, the present section reviews some necessary material on operator theory that will be needed in Section \ref{sec:POCSlim} to prove the above mentioned POCS behavior.

\subsection{Sampling operator $S$}

In finite dimension, the above mentioned transformation $S$  is typically formalized as a matrix. One would then naturally estimate $x$ by applying the matrix pseudo-inverse  $S^+$ on $\vs$. Under certain conditions, this is also possible in infinite dimension, but the setting is substantially more involved. In this case, a linear transformation typically involves infinite summations whose convergence must be guaranteed with respect to some norm. With the needed tool of orthogonal projection and the objective to invert $S$, this requires both the domain and the destination of $S$ to be Hilbert spaces. A rigorous way to construct our operator $S$ is to present it as follows:
\begin{equation}\label{S}
S:\begin{array}[t]{rcl}
 \scA & \rightarrow &\scD\\
 u & \mapsto & Su:=\big(\langle u, g_k\rangle\big)\inZ
 \end{array}
 \end{equation}
where
\begin{equation}\label{D}
\scD:=\Big\{\vc:\big(\c_k/\|g_k\|\big)\inZ\in\ell^2(\Z)\Big\}
\end{equation}
and $\ell^2(\Z)$ is the set of square-summable sequences indexed by $\Z$. We call $S$ the {\em sampling operator}. While $\scA$ is by default seen as a Hilbert space with respect to the inner product $\langle\cdot,\cdot\rangle$ of the ambient space $\scH$, $\scD$ is a Hilbert space with respect to the inner product $\langle\cdot,\cdot\rangle_\scD$ defined by
\begin{equation}\label{innerD}
\forall \vc,\vd\in\scD,\qquad\langle\vc,\vd\rangle_\scD:=\smallsum{k\in\Z}\midfrac{\c_k\,\d_k}{\|g_k\|^2}.
\end{equation}
Next, note that $Su$ does belong to $\scD$ for all $u\in\scA$. Writing $Su=(\c_k)\inZ$, this is because  $(\c_k/\|g_k\|)\inZ=\big(\big\langle u,g_k/\|g_k\|\big\rangle\big)\inZ$, which belongs to $\ell^2(\Z)$ since  $(g_k/\|g_k\|)\inZ$ is  orthonormal. When $\Z$ is finite, note also that $\ell^2(\Z)$ is just $\RR^\Z$. Finally, when the sampling is uniform (implying that $\|g_k\|$ is constant), note that $\langle\cdot,\cdot\rangle_\scD$ coincides with the canonical inner product $\langle\cdot,\cdot\rangle_2$ of $\ell^2(\Z)$, up to a scaling factor. Otherwise, $\langle\cdot,\cdot\rangle_\scD$ can be interpreted as a weighted version of $\langle\cdot,\cdot\rangle_2$ to compensate for the non-uniformity.

The sampling operator leads to a more concise way to characterize the set of consistent estimates $\scA_\vs$. Indeed, it follows from \eqref{As} and \eqref{S} that
\begin{equation}\label{As2}
\scA_\vs=\big\{v\in\scA:Sv=\vs\big\}=S^{-1}(\vs).
\end{equation}
With this presentation,
\begin{equation}\label{As-empty}
\scA_\vs\neq\emptyset\qquad\Leftrightarrow\qquad\vs\in\ran(S)
\end{equation}
where $\ran(S)$ denotes the range $S$. Finally, it will be useful to note that the null space $\null(S)$ of $S$ is given by
\begin{equation}\label{nullS}
\null(S)=\scF^\perp
\end{equation}
as a result of \eqref{F} and \eqref{inner-proj}.

\subsection{Adjoint operator $S^*$}

In finite dimension, the matrix pseudo-inverse $S^+$ is fundamentally linked to the matrix transpose $S^\top$ of $S$. Generalizing pseudo-inversion in infinite dimension will also be based on a generalization of matrix transpose, which is as follows: a linear operator $S^*$ from  $\scD$ back to $\scA$ is said to be adjoint to $S$ when
\begin{equation}\label{adjoint-eq}
\forall u\in\scA,~\forall\vc\in\scD,\qquad
\langle u,S^*\vc\rangle=\langle  S u,\vc\rangle_\scD.
\end{equation}
A difficulty is that this uniquely defines $S^*$ only when $S$ is bounded, i.e., when there exists $\beta\geq0$ such that
\begin{equation}\label{boundedS}
\forall u\in\scA,\qquad\|Su\|_\scD\leq\beta\|u\|
\end{equation}
where $\|\cdot\|_\scD$ is the norm induced in $\scD$ by $\langle\cdot,\cdot\rangle_\scD$. Let us verify that this is indeed realized with the operator $S$ of \eqref{S}. As
\begin{equation}\label{normD}
\forall \vc\in\scD,\qquad\|\vc\|_\scD^2=\langle\vc,\vc\rangle_\scD=\smallsum{k\in\Z}\midfrac{|\c_k|^2}{\|g_k\|^2}
\end{equation}
from \eqref{innerD},
\begin{equation}\label{Su-norm}
\|Su\|_\scD^2=
\smallsum{k\in\Z}\midfrac{|\langle u, g_k\rangle|^2}{\|g_k\|^2}=
\smallsum{k\in\Z}\big|\big\langle u,\midfrac{g_k}{\|g_k\|}\big\rangle\big|^2=\|P_\scG u\|^2
\end{equation}
from \eqref{PL} and by orthonormality of $(g_k/\|g_k\|)\inZ$. By Bessel's inequality, $P_\scG$ is non-expansive. So \eqref{boundedS} is satisfied with $\beta=1$. The next expression gives the explicit description of $S^*$.
\line
\begin{proposition}
The adjoint $S^*$ to the operator $S$ of \eqref{S} is given by
\begin{equation}\label{S*}
 S^*:\begin{array}[t]{rcl}
\,\scD & \rightarrow &\scA\\
\vc=(\c_k)\inZ&\mapsto&S^*\vc=\ssum{k\in \Z}\midfrac{\c_k}{\|g_k\|^2}\, {\widetilde g}_k
 \end{array}.
 \end{equation}
\end{proposition}

\begin{IEEEproof}
Using \eqref{S*} as the definition of $S^*$, we have for all  $u\in\scA$ and $\vs\in\scD$,
\begin{equation}\nonumber
\langle u,S^*\vc\rangle=
\Big\langle u,\smallsum{k\in\Z}\midfrac{\c_k}{\|g_k\|^2} {\widetilde g}_k\Big\rangle=
\smallsum{k\in\Z}\midfrac{\langle u, {\widetilde g}_k\rangle\c_k}{\|g_k\|^2}=
\langle  S u,\vc\rangle_\scD
\end{equation}
as a result of \eqref{inner-proj} and \eqref{innerD}.
\end{IEEEproof}
\ppnoi
The well known basic properties of operator adjoint \cite{Luenberger69} are
\begin{equation}\label{adjoint-prop}
\null(S)=\ran(S^*)^\perp\quad\mbox{and}\quad\null(S^*)=\ran(S)^\perp.
\end{equation}

\subsection{Pseudo-inverse $S^\dagger$}\label{subsec:pseudo}

The Moore-Penrose pseudo-inverse of $S$ is the linear operator $S^\dagger$ from $\scD$ back to $\scA$ such that \cite[\S11]{wang2018generalized}
\begin{eqnarray}
&\forall\vs\in\scD,\qquad S^\dagger\vs=\argmin\limits_{v\in\scM_\vs}\|v\|,\qquad\label{S+}\\
\mbox{where}&&\quad\quad\nonumber\\
&\scM_\vs:=\big\{v\in\scA:\| S v-\vs\|_\scD\mbox{ is minimized}\big\}.\label{Ms}
\end{eqnarray}
Contrary to the finite dimensional case, $S^\dagger$ does not always exist for the simple reason that $\| S v-\vs\|_\scD$ does not always have a minimizer $v\in\scA$. This minimizer systematically exists only when
\begin{equation}\label{closed-range}
\ran(S)\mbox{ is closed in }\scD.
\end{equation}
Assuming this condition, the following are some useful properties of $S^\dagger$ that can be found in \cite[\S11]{wang2018generalized}:
\begin{subequations}\label{pseudoprop}
\begin{align}
S^\dagger S&=P_{\ran(S^*)},\label{pseudoprop0}\\
S\,S^\dagger&=P_{\ran(S)},\label{pseudopropa}\\
\ran(S^\dagger)&=\ran(S^*),\label{pseudopropb}\\
\null(S^\dagger)&=\null(S^*)\label{pseudopropc}
\end{align}
\end{subequations}
(\eqref{pseudopropc} is a consequence of Theorem 11.1.6 of \cite{wang2018generalized}). By assumption of \eqref{closed-range}, note that $\ran(S^*)$ is also closed (see for example Lemma 2.5.2 of \cite{Christensen08}). This allows the valid statement of \eqref{pseudoprop0} and implies that
\begin{equation}\label{ranS*}
\ran(S^\dagger)=\ran(S^*)=\null(S)^\perp=\scF
\end{equation}
as a result of \eqref{pseudopropb}, \eqref{adjoint-prop}, \eqref{nullS} and the fact that $\scF$ is closed.
Next is a result on the algebraic structure of the set $\scM_\vs$ of \eqref{Ms}.
\line
\begin{proposition}\label{prop:Ms}
Assuming that $\ran(S)$ is closed,
\begin{equation}\label{bAs}
\forall\vs\in\scD,\qquad
\scM_\vs=\scA_\bvs=S^\dagger\vs+\scF^\perp
\end{equation}
where $\bvs:=P_{\ran(S)}\vs$.
\end{proposition}
\line
\begin{IEEEproof}
Let $\vs\in\scD$. According to Theorem 11.1.1 of \cite{wang2018generalized}, $\| S v-\vs\|_\scD$ is minimized if and only if $Sv=\bvs$. This proves that $\scM_\vs=\scA_\bvs$. It follows from \eqref{pseudopropa} that $S(S^\dagger\vs)=P_{\ran(S)}\vs=\bvs$. It then results from \eqref{As2} that  $S^\dagger\vs\in\scA_\bvs$. The second equality of \eqref{bAs} is then obtained by applying \eqref{As-set} after replacing $\vs$ and $u$ by $\bvs$ and $S^\dagger\vs$, respectively.
\end{IEEEproof}
\ppnoi
Remark: When $\vs\in\ran(S)$, one simply has $\scM_\vs=\scA_\vs$ since $\bvs=\vs$ in this case. Now, as a result of \eqref{As-empty}, while $\scA_\vs$ becomes empty when $\vs\notin\ran(S)$, note that $\scM_\vs=\scA_\bvs$ is never empty since $\bvs\in\ran(S)$.

\subsection{Stable sampling}

While condition \eqref{closed-range} may seem abstract, it is seen in this section as a necessary condition for stable reconstruction. The next proposition first gives some equivalent analytical properties to the fact that $\ran(S)$ is closed. One of them involves the {\em reduced minimum modulus} of $S$, which is defined by
\begin{equation}\label{modulus}
\gamma(S):=\inf_{u\in\scF\backslash\{0\}}\midfrac{\|Su\|_\scD}{\|u\|~}
\end{equation}
given that $\scF=\null(S)^\perp$ from \eqref{nullS} as $\scF$ is closed.
\line
\begin{proposition}\label{prop:equiv1}
The following statements are equivalent:
\begin{enumerate}[label=(\roman*)]
\item $\ran(S)$ is closed in $\scD$.
\item $\gamma(S)>0$,
\item $(f_k)\inZ$ is a {\em frame} of $\scF$ with
$f_k:=\widetilde g_k/\|g_k\|$, $k\in\Z$.
\end{enumerate}
\end{proposition}
\ppnoi
The definition of a frame and the proof of this result are in Appendix \ref{app:equiv1}. The connection of (ii) with reconstruction stability is as follows. Consider the estimation of $x\in\scA$ from noisy samples $\vs=Sx+\ve$ where $\ve$ is some error sequence. To simplify the problem, assume that $\scF=\scA$, guaranteeing uniqueness of reconstruction, and $\ve\in\ran(S)$. Then $Su=\vs$ will imply that $u=x+e$ where $e$ is in $\scF$ while satisfying $Se=\ve$. As $\|e\|/\|\ve\|_\scD=\|e\|/\|Se\|_\scD$, this ratio has no upper bound if (ii) is not satisfied. In other words, for a given sampling error norm $\|\ve\|_\scD$, the reconstruction error norm $\|e\|$ can be arbitrarily large. This prevents reconstruction stability. On the other hand, if (ii) is satisfied, we will automatically have $\|e\|\leq\|\ve\|_\scD/\alpha$. For this reason, (ii) is called the condition of ``stable sampling''\footnote
{In the literature, ``stable sampling'' also implies uniqueness of reconstruction, .i.e., $\scF=\scA$. But in this paper, we use this expression in the more general context where $\scF\subset\scA$. This refers more generally to the stability of reconstruction of a consistent estimate.} \cite{Benedetto92,eldar2015sampling}. Note that this is indeed an intrinsic property of the sampling (and not of the chosen reconstruction method) since it is solely dependent on the sampling kernels $(g_k)\inZ$ as seen for example in the equivalent formulation of (iii).

As a final remark,  $\ran(S)$ is always closed when the number of samples is finite, as is always the case in practice. This is because $\ran(S)$ is of finite dimension. At the same time, (ii) is systematically satisfied in this case with $\alpha$ equal to the smallest positive singular value of $S$.

\section{Connection of POCS iteration to sampling pseudo-inversion}\label{sec:POCSlim}

In this section, we show that the POCS iteration $u\up{n+1}=P_\scA P^{\lambda_n}_{\scH_\vs} u\up{n}$ of \eqref{x-iter3} belongs to a larger family of algorithms that lead to the pseudo-inversion of $S$, and that is simultaneously a generalization of the {\em frame algorithm} \cite{Duffin52,Benedetto92,Feichtinger94}. At the end of the section, this will allow us to deduce the theoretical effect of sampling noise on the limit of $u\up{n}$ .

\subsection{Generic form of iterated map}\label{subsec:gen-form}

The first step is to connect $P_\scA P^\lambda_{\scH_\vs}$ to the operators $S$ and $S^*$. It follows from \eqref{eq:proj} and \eqref{relaxed-P} that
\begin{equation}\label{PCs-relax}
\forall u\in\scH,\qquad P^\lambda_{\scH_\vs}u=u+\lambda P_\scG(x-u).
\end{equation}
We then obtain from \eqref{Rsl} and \eqref{PL} that
\begin{equation}\nonumber
\forall u\in\scA,\qquad P_\scA P^\lambda_{\scH_\vs}u=u+\lambda\smallsum{k\in\Z}\frac{\s_k-\langle u,g_k\rangle}{\| g_k\|^2}\,\widetilde g_k
\end{equation}
where $\widetilde g_k$ was defined in \eqref{tgk}.
Since $\vs-Su=\big(\s_k{-}\langle u,g_k\rangle\big)\inZ$, it then follows from  \eqref{S*} that
\begin{equation}\label{PAPCs2a}
\forall u\in\scA,\qquad P_\scA P^\lambda_{\scH_\vs}u=u+\lambda S^*\!(\vs-Su).
\end{equation}

\subsection{Generalized frame algorithm}\label{subsec:gen-frame}

The previous section has shown that the iteration of \eqref{x-iter3} is of the type
\begin{eqnarray}
&u\up{n+1}=R^{\lambda_n}_\vs u\up{n},\quad n\geq0\quad\label{x-iter4}\\[0.5ex]
\mbox{where}\qquad\quad&R^\lambda_\vs u:=u+\lambda S^*\!(\vs-Su)&\qquad\qquad\quad~~\label{Rsl}
\end{eqnarray}
for any $u\in\scA$ and $\lambda\in\RR$. We are going to see a direct connection between the iteration of \eqref{x-iter4} and the pseudo-inversion of $S$ under very general assumptions on $S$. Specifically, we will only assume in this subsection and the next that $S$ is a bounded operator from $(\scA,\langle\cdot,\cdot\rangle)$ to any other Hilbert space $(\scD,\langle\cdot,\cdot\rangle)_\scD$ (not necessarily made of sequences). Section \ref{sec:pseudo} will still be applicable, except (\ref{S}-\ref{innerD}),  (\ref{normD}-\ref{S*}) and Proposition \ref{prop:equiv1}\,(iii) (\eqref{ranS*} will be used as the definition of $\scF$).

Starting from $u\up{0}\in\scF$, note first from \eqref{x-iter4} and \eqref{Rsl} that $u\up{n}$ must remain in $\scF$ since $\ran(S^*)=\scF$ from \eqref{ranS*}. Therefore, if $u\up{n}$ is convergent, its limit must be a fixed point of $R^\lambda_\vs$ in $\scF$. The next proposition then gives an immediate connection of \eqref{x-iter4} with the pseudo-inversion of $S$.
\line
\begin{proposition}\label{prop:fixed-point}
Assume that $S$ has closed range. For any $\vs\in\scD$ and $\lambda\in\RR$, $S^\dagger\vs$ is a fixed point of $R^\lambda_\vs$ in $\scF$.
\end{proposition}
\line
\begin{IEEEproof}
Note first that $S^\dagger\vs\in\scF$ from \eqref{ranS*}. Next, it is easy to see from \eqref{Rsl} that any solution $u$ to the equation $S^*(\vs-Su)=0$ is a fixed point of $R^\lambda_\vs$. It follows from  \eqref{pseudopropa} that $\vs-SS^\dagger\vs=\vs-P_{\ran(S)}\vs)\in\ran(S)^\perp=\null(S^*)$ due to \eqref{adjoint-prop}. So $S^*(\vs-SS^\dagger\vs)=0$.
\end{IEEEproof}
\ppnoi
What remains is to find a condition for \eqref{x-iter4} to be convergent. This question turns out to be resolved in the analysis of the {\em frame algorithm}  \cite{Duffin52,Benedetto92,Feichtinger94} which falls in the particular case where
$\scD=\ell^2(\ZZ)$, $\vs\in\ran(S)$ and $\scF=\scA$.
In the following proposition, we reproduce this analysis under the general Hilbert space assumptions of this section.
\line
\begin{proposition}\label{prop:frame-alg}
For any $\vs\in\scD$ and $\lambda\in\RR$,
\begin{eqnarray}
&\forall v,w\in\scF,\qquad\|R^\lambda_\vs v-R^\lambda_\vs w\|\leq c_\lambda\,\|v-w\|\label{contract}\\[1ex]
\lefteqn{\mbox{where}}\qquad~\,&
c_\lambda=\max\big(|1{-}\lambda\|S\|^2|,|1{-}\lambda\,\gamma(S)^2|\big),&\qquad~\,\label{contract-coef}\\[0.5ex]
&\|S\|:=\displaystyle\sup_{u\in\scF\backslash\{0\}}\|Su\|_\scD/\|u\|\label{op-norm}
\end{eqnarray}
and $\gamma(S)$ is defined in \eqref{modulus}.
\end{proposition}
\line
\begin{IEEEproof}
It follows from \eqref{Rsl} that
\begin{equation}\label{RQ}
R^\lambda_\vs u=Q^\lambda u+\lambda S^*\vs\quad\mbox{where}\quad Q^\lambda u:=u-\lambda S^*Su.
\end{equation}
Thus, $R^\lambda_\vs v-R^\lambda_\vs w=Q^\lambda(v-w)$. Then, \eqref{contract} is satisfied with $c_\lambda:=\sup_{u\in\scF\backslash\{0\}}\|Q^\lambda u\|/\|u\|$. Because $Q^\lambda$ is self-adjoint, it is known (see for example \cite[\S2.13]{Conway90}) that $c_\lambda$ is equivalently the supremum of $\big|\langle u,Q^\lambda u\rangle\big|$ over the set $U_\scF:=\{u\in\scF:\|u\|=1\}$.
We have
$$\langle u,Q^\lambda u\rangle=\|u\|^2-\lambda\|Su\|_\scD^2$$
since $\langle u,Q^\lambda u\rangle=\langle u, u\rangle-\lambda\langle u, S^*Su\rangle=\|u\|^2-\lambda\langle Su,Su\rangle_\scD$ due to \eqref{adjoint-eq}. It then follows from \eqref{op-norm} and \eqref{modulus} that the infimum and the supremum of $\langle u,Q^\lambda u\rangle$ over  $U_\scF$ are $1-\lambda\|S\|^2$ and $1-\lambda\,\gamma(S)^2$, respectively. This leads to \eqref{contract-coef}.
\end{IEEEproof}
\ppnoi
If $c_\lambda<1$ in \eqref{contract} for some given $\lambda$, then $R^\lambda_\vs$ is a contraction within $\scF$ and $u\up{n}$ from \eqref{x-iter4} is guaranteed to converge starting from $u\up{0}\in\scF$ with $\lambda_n=\lambda$ for all $n\geq0$. By property of contraction \cite{smart1980fixed}, $S^\dagger\vs$ will have to be the unique fixed point of $R^\lambda_\vs$ in $\scF$ and hence the limit of $u\up{n}$. When $S$ is bounded of closed range, it is shown in \cite{Duffin52,Benedetto92} that $c_\lambda$ is minimized with $\lambda=2/(\gamma(S)^2{+}\|S\|^2)$ of minimum value $(\|S\|^2{-}\gamma(S)^2)/(\|S\|^2{+}\gamma(S)^2)$. The interest of the present paper will be more generally to know the values of $\lambda$ that lead to a coefficient $c_\lambda<1$.

\begin{proposition}\label{prop:lambda-cond}
Let $c_\lambda$ be defined by \eqref{contract-coef}. For $\eps>0$,
\begin{equation}\label{eps-contract}
\lambda\in\big[\eps,2\|S\|^{-2}{-}\eps\big]\quad\Rightarrow\quad c_\lambda\leq\rho_\eps
:=1-\eps\,\gamma(S)^2.
\end{equation}
If $S$ has closed range, $\rho_\eps<1$.
\end{proposition}
\line
\begin{IEEEproof}
By construction, $\gamma(S)\leq\|S\|$. In this case, it can be verified that $c_\lambda$ is more specifically equal to $\max\big(\lambda\|S\|^2{-}1\,,\,1{-}\lambda\,\gamma(S)^2\big)$. Now, while $\lambda\|S\|^2{-}1\leq(2\|S\|^{-2}{-}\eps)\|S\|^2{-}1=1-\eps\|S\|^2$, we have $1{-}\lambda\,\gamma(S)^2\leq1{-}\eps\,\gamma(S)^2$. The second upper bound is the larger one, which proves \eqref{eps-contract}.
When $S$ has closed range, $\gamma(S)>0$ by Proposition \ref{prop:equiv1}, and hence $\rho_\eps<1$.
\end{IEEEproof}

\subsection{Generalized iteration}

In the previous section, we focused on the case where $u\up{0}\in\scF$ and $(\lambda_n)_{n\geq0}$ is constant. We now give a general result of convergence of $u\up{n}$ from \eqref{x-iter4} in absence of these two conditions. The case where $u\up{0}\notin\scF$ will be resolved thanks to the following property.
\line
\begin{proposition}\label{prop:out-of-F}
For all $v\in\scF$ and $w\in\scF^\perp$,
$$R^\lambda_\vs(v+w)=R^\lambda_\vs v+w.$$
\end{proposition}

\begin{IEEEproof}
 It follows from \eqref{RQ} that $R^\lambda_\vs(v{+}w)=Q^\lambda(v{+}w)+\lambda S^*\vs=R^\lambda_\vs v+Q^\lambda w$ by linearity of $Q^\lambda$. But due to \eqref{nullS}, $Sw=0$. Therefore, $Q^\lambda w=w$.
\end{IEEEproof}
\ppnoi
By combining this with Propositions \ref{prop:fixed-point}-\ref{prop:lambda-cond}, we obtain the following result.
\line
\begin{theorem}\label{theo:genlim}
Assuming that $S$ is bounded of closed range, let $(u\up{n})_{n\geq0}$ be a sequence satisfying the recursion of \eqref{x-iter4} for some $\vs\in\scD$ and initial iterate $u\up{0}\in\scA$, with a sequence of coefficients $(\lambda_n)_{n\geq0}$ such that $\lambda_n\in\big[\eps,2\|S\|^{-2}{-}\eps\big]$ for all $n\geq0$ and some constant $\eps>0$. Then, $(u\up{n})_{n\geq0}$ is convergent of limit
\begin{eqnarray}
&u\up{\infty}=S^\dagger\,\vs+P_{\scF^\perp}u\up{0}=P_{\scM_\vs}u\up{0}\label{x-inf3}\\[0.5ex]
\mbox{and}\quad&\|u\up{n}\!-u\up{\infty}\|\leq\rho_\eps^n\|u\up{0}\!-u\up{\infty}\|,\quad\forall n\geq0&\quad\quad\label{contract2}
\end{eqnarray}
where $\scM_\vs$ and $\rho_\eps$ are defined in \eqref{Ms} and \eqref{eps-contract}.
\end{theorem}
\line
\begin{IEEEproof}
Let $u\up{0}\in\scA$. We can write that $u\up{0}=v\up{0}+w\up{0}$ where $v\up{0}:=P_\scF u\up{0}$ and $w\up{0}:=P_{\scF^\perp}u\up{0}$. Let $v\up{n+1}:=R^{\lambda_n}_\vs v\up{n}$ for all $n\geq0$. Since $v\up{0}\in\scF$, we saw in the previous section that $v\up{n}$ remains in $\scF$ for all $n\geq0$. By assumption on $\lambda_n$, it follows from Proposition \ref{prop:lambda-cond} that $c_{\lambda_n}\leq\rho_\eps$. Then, by applying this with Propositions \ref{prop:fixed-point} and \ref{prop:frame-alg} on the sequence $(v\up{n})_{n\geq0}$, we obtain that $\|v\up{n+1}\!-S^\dagger\vs\|=\|R^{\lambda_n}_\vs v\up{n}\!-R^{\lambda_n}_\vs(S^\dagger\vs)\|\leq\rho_\eps\|v\up{n}\!-S^\dagger\vs\|$. Since $\rho_\eps<1$, this proves that $v\up{n}$ is convergent of limit $v\up{\infty}=S^\dagger\vs$ and that
\begin{equation}\label{v-ineq}
\|v\up{n}\!-v\up{\infty}\|\leq\rho_\eps^n\|v\up{0}\!-v\up{\infty}\|,\quad\forall n\geq0.
\end{equation}
Meanwhile, one easily finds by induction from Proposition \ref{prop:out-of-F} that $u\up{n}=v\up{n}+w\up{0}$ for all $n\geq0$. Therefore, $u\up{n}$ tends to $u\up{\infty}=v\up{\infty}+w\up{0}=S^\dagger\,\vs+P_{\scF^\perp}u\up{0}$. On the one hand, \eqref{contract2} is deduced from \eqref{v-ineq} by a mere space translation by $w\up{0}$. On the other hand, it results from \eqref{bAs} that
$P_{\scM_\vs}u\up{0}=P_{(\scF^\perp+S^\dagger\vs)}u\up{0}=P_{\scF^\perp}(u\up{0}\!-S^\dagger\vs)+S^\dagger\vs=
P_{\scF^\perp}u\up{0}+S^\dagger\vs$ since $S^\dagger\vs\in\scF$ from \eqref{ranS*}.  This proves \eqref{x-inf3}.
\end{IEEEproof}

\subsection{Application to POCS iteration}

A known shortcoming of the frame algorithm is that the range of admissible relaxation coefficients depends on  $\|S\|$ which is not necessarily accessible in practice. This problem is however less critical with the iteration of \eqref{x-iter3} which corresponds to $R^\lambda_\vs=P_\scA P^\lambda_{\scH_\vs}$. Returning to the beginning of Section \ref{sec:POCSlim}, this is the case of \eqref{Rsl} where $S$ is defined by \eqref{S} with \eqref{D} under the assumption of \eqref{cond0}. Because of \eqref{Su-norm} and the fact that $P_\scG$ is non-expansive, we know that $\|S\|\leq1$. In this case, the interval $\big[\eps,2\|S\|^{-2}{-}\eps\big]$ includes $[\eps,2{-}\eps]$ as a subset. Theorem \ref{theo:genlim} then has the  following consequence.
\line
\begin{corollary}
Let $(u\up{n})_{n\geq0}$ be recursively defined by \eqref{x-iter3} for some $\vs\in\scD$, coefficients $(\lambda_n)_{n\geq0}$ and initial iterate $u\up{0}\in\scA$ with the following assumptions: the operator $S$ defined by \eqref{S} and \eqref{D}  has closed range and $\lambda_n\in[\eps,2{-}\eps]$ for all $n\geq0$ and some $\eps>0$. Then $(u\up{n})_{n\geq0}$ is convergent of limit $u\up{\infty}$ satisfying \eqref{x-inf3} and \eqref{contract2}.
\end{corollary}
\ppnoi
Note that the above assumption on $\lambda_n$ implies that $\sum_{n\geq0}\lambda_n(2{-}\lambda_n)=+\infty$. Theorem \ref{theo:relax} is then applicable {\em without} the condition that $\ran(S)$ is closed, and hence with no stable sampling condition. The requirement of this theorem, however, is that $\scA_\vs$ must be non-empty. Now when $\ran(S)$ is closed,  \eqref{x-inf3} can be seen as an extension of \eqref{x-inf2} allowing $\scA_\vs$ to be empty (due to sampling noise for example) following the remark at the end of Section \ref{subsec:pseudo}. Also, we have from \eqref{contract2} that the convergence of $u\up{n}$ is {\em linear} (referring to the exponent of $\gamma^n$), which is not guaranteed by Theorem \ref{theo:relax}.

\subsection{Consequence on noisy sampling}\label{subsec:noise}

In absence of sampling noise. we already know the signal implications of the POCS limit $P_{\scA_\vs} u\up{0}$ of \eqref{x-inf2}. In practice however, the sampling sequence that is injected into \eqref{x-inf2} is often not $\vs$, but
\begin{equation}\label{samp-err}
\hat\vs:=\vs+\ve\qquad\mbox{where}\qquad\vs:=Sx
\end{equation}
and $\ve=(\e_k)\inZ$ is some unknown error sequence.
It then follows from \eqref{x-inf3} and \eqref{bAs} that
\begin{align*}
u\up{\infty}&=P_{\scF^\perp}u\up{0}\!+ S^\dagger\hat\vs= \big(P_{\scF^\perp}u\up{0}\!+ S^\dagger\vs\big)+  S^\dagger\ve\\
&=P_{\scA_\vs} u\up{0}+S^\dagger\ve
\end{align*}
since $\vs\in\ran(S)$ and hence $\bvs=\vs$.
While $P_{\scA_\vs} u\up{0}$ is the noise-free POCS iteration limit,  $S^\dagger\ve$ is the deviation of this limit due to the sampling error sequence $\ve$.
\line
\begin{proposition}
Let $\bve:=P_{\ran( S)}\ve$.  Then, $$S^\dagger\ve=S^\dagger\bve\qquad\mbox{and}\qquad\|\bve\|_\scD\leq\|\ve\|_\scD$$
\end{proposition}

\begin{IEEEproof}
It follows from \eqref{pseudopropc} and \eqref{adjoint-prop} that $\null(S^\dagger)=\null(S^*)=\ran(S)^\perp$.
By construction, $\bve-\ve\in\ran(S)^\perp=\null(S^\dagger)$. Thus $S^\dagger\ve=S^\dagger\bve$. Meanwhile, $\|\bve\|_\scD\leq\|\ve\|_\scD$ is from Bessel's inequality.
\end{IEEEproof}
\ppnoi
Hence, only the component $\bve$ of $\ve$ in $\ran(S)$ contributes to the reconstruction deviation. Thus, the POCS iteration has a filtering effect on the noise sequence $\ve$. Meanwhile, the error $\bve$ is irreversible. Indeed, since $\bve\in\ran(S)$, then $\bve=S\bar e$ for some $\bar e\in\scA$. There is no more knowledge to discriminate $x$ from $x+\bar e$.

To have a strict inequality $\|\bve\|_\scD<\|\ve\|_\scD$, note that $\ran(S)$ must be a proper subspace of $\scD$. This necessitates some oversampling. The higher the oversampling ratio is, the smaller $\ran(S)$ is compared to $\scD$, and the smaller $\|\bve\|_\scD$ is compared to $\|\ve\|_\scD$. This corresponds to the noise-shaping effect of oversampling in uniform sampling \cite{norsworthy1997delta}.

\section{Practical implementation aspects}\label{sec:bin}

We present a rigorous way to implement the iterative part of \eqref{x-iter4} in discrete-time, without the need to involve generic discrete-time decompositions of the signals of $\scA$ (such as sinc-basis decompositions for bandlimited signals). Although not studied in this paper, we will then touch on the issue of finite-complexity implementations.

\subsection{Discrete-time implementation of iteration}\label{subsec:discrete}

Using \eqref{PAPCs2a}, \eqref{x-iter4} takes the form
\begin{equation}\label{x-iter-relax}
u\up{n+1}=u\up{n}+\lambda_n S^*\!(\vs- S u\up{n}),\qquad n\geq0.
\end{equation}
This is in practice an iteration of continuous-time functions. For digital signal processing, one expects this iteration to be discretized. Now, whether the space $\scA$ has a countable basis or not, there is a way to obtain $u\up{n}$ by a pure discrete-time iteration in $\scD$. The principle is as follows. Note from \eqref{x-iter-relax} that $u\up{n+1}-u\up{n}\in\ran(S^*)$ for all $n\geq0$. Hence, $u\up{n}-u\up{0}$ must be in $\ran(S^*)$ as well. This implies that there exists some discrete-time sequence $\vc\up{n}\in\scD$ such that $u\up{n}=u\up{0}+S^*\vc\up{n}$. The next proposition implies a way to construct $\vc\up{n}$ recursively.
\line
\begin{proposition}\label{prop:sys-equiv}
For any given initial estimate $u\up{0}$, the iterate $u\up{n}$ of \eqref{x-iter4} is equivalently obtained by iterating the system
\begin{subequations}\label{sys}
\begin{align}
\vc\up{n+1}&=\vc\up{n}+ \lambda_n(\vs_0- S S^*\vc\up{n})\label{discr-iter}\\
u\up{n}&=u\up{0}+S^*\vc\up{n}\label{xc}
\end{align}
\end{subequations}
for $n\geq0$, where $\vs_0:=\vs-S u\up{0}$, starting with $\vc\up{0}=\bzero$.
\end{proposition}
\line
\begin{IEEEproof}
It follows from \eqref{xc} and \eqref{discr-iter} that
\begin{align*}
u\up{n+1}&=u\up{0}\!+ S^*\vc\up{n+1}=u\up{0}\!+ S^*\big(\vc\up{n}{+}\lambda_n(\vs_0{-}S S^*\vc\up{n})\big)\\
&=\big(u\up{0}\!+S^*\vc\up{n}\big)+\lambda_n S^*\!\big(\vs-Su\up{0}\!-S S^*\vc\up{n}\big)\\
&=u\up{n}+\lambda_n S^*\!(\vs- S u\up{n})
\end{align*}
which leads to \eqref{x-iter-relax}.
\end{IEEEproof}
\ppnoi
The outstanding contribution of \eqref{sys} compared to \eqref{x-iter4} is that the pure discrete-time operation of \eqref{discr-iter} can be iterated {\em alone} until the targeted iteration number $m$. Then, the continuous-time operation of \eqref{xc} just needs to be executed {\em once} at $n=m$. The operator $S S^*$ involved in \eqref{discr-iter} can be seen as a square matrix of coefficients
\begin{equation}\label{inner-mat}
 S S^*=\begin{bmatrix}\displaystyle\frac{\langle {\widetilde g}_{k'},g_k\rangle}{\|g_{k'}\|^2}\end{bmatrix}_{(k{,}k')\in\Z\times\Z}
\end{equation}
which needs to be predetermined before the iteration. In general, the coefficients can only be obtained numerically. It was proposed in \cite{Thao21a,Thao22a} to obtain them from a {\em single-argument} lookup table.

\subsection{Finite-complexity implementation}

A remaining issue is the finite-complexity implementation of \eqref{sys}. The most critical part is the recurrent operation of $SS^*$ in \eqref{discr-iter}. For any $\vc=(\s_k)\inZ\in\scD$, the $k$th component of $SS^*\vc$ is
\begin{equation}\label{SS*}
\left.SS^*\vc\right|_k=\smallsum{k'\in\Z}\h_{k,k'}\,\s_{k'}
\end{equation}
where $\h_{k,k'}:=\langle {\widetilde g}_{k'},g_k\rangle_2/\|g_{k'\!}\|_2^2$. For concrete analysis, let us assume the traditional case where $\scH=L^2(\RR)$, $\scA=\scB$ and $\Z\subset\ZZ$. When the sampling is uniform, i.e., $g_k(t)=g(t{-}kT)$ for some $g(t)\in L^2(\RR)$, $\h_{k,k'}$ reduces to $\h_{k'-k}$ where $\h_k:=\langle\widetilde g,g_k\rangle_2/\|g\|_2^2$. So $SS^*\vc$ is just a convolution operation. Note that $(\h_k)_{k\in\ZZ}$ are simply the squared uniform samples of $\widetilde g(t)$ with some scaling factor. With the bandlimitation, $\h_k$ decays towards infinity in a sinc-like manner. For operation of finite complexity, it is necessary to approximate $\h_k$ using standard FIR windowing methods, which creates reconstruction distortions. When the sampling is nonuniform, $SS^*$ loses its time invariance. But it can still be seen as a linear filter of time-varying impulse response $\h_{k,k'}$, with expected decays when $k'$ gets far away from $k$.
However, the windowing of time-varying filters plus the effect of filter distortions to the POCS iteration remain virgin topics that require new substantial investigations not tackled in this article and to be addressed in the future. Some preliminary experiments can been found in \cite{Thao21a}.

\section{Multi-channel orthogonal sampling}\label{sec:multi-TEM}

\begin{figure}
\centerline{\hbox{\scalebox{0.9}{\includegraphics{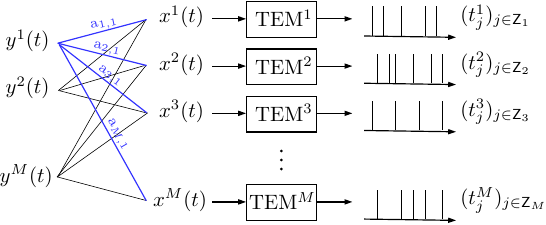}}}}
\caption{Multi-channel system of  time-encoding machines (TEM) from \cite{Adam20,Adam21}.}\label{fig:Karen}
\end{figure}

In this section, we illustrate the theoretical power of our formalism by revisiting the sampling/reconstruction system designed in \cite{Adam20,Adam21} for multi-channel time encoding. A basic POCS algorithm was used to reconstruct a multi-channel signal from the elaborate sampling system shown in Fig. \ref{fig:Karen}\footnote{The letters `$x$' and `$y$' from \cite{Adam20,Adam21} have been interchanged in Fig. \ref{fig:Karen} to be compatible with the notation of the present article.}. We show that this encoding system turns out to satisfy the abstract conditions of \eqref{sample0} and \eqref{cond0}. While the reconstruction iterates of \cite{Adam20,Adam21} coincide with those of \eqref{x-iter2} under our formalism, our theory uncovers the full pseudo-inversion property of this method, which was only studied in a noise-free case of perfect reconstruction in these references. Moreover, while the reconstruction method was mostly presented at a conceptual level in  \cite{Adam20,Adam21}, our abstract reformulation simultaneously allows a more explicit descriptions of practical implementations.

\subsection{System description}

The time-encoding system of \cite{Adam20,Adam21}  assumes that the source signals are multidimensional bandlimited functions
$$\by(t)=(y^1(t),{\cdots},y^N(t))\in\scB^N.$$
Next, instead of sampling the functions $y^i(t)$ individually, the system first expands $\by(t)$ into a redundant representation
$$\bx(t):=\A\by(t),\qquad t\in\RR$$
where $\A$ is an $M{\times}N$ matrix that is assumed in \cite{Adam20,Adam21} to be full rank with $M\geq N$. In our present analysis, we will not necessarily assume so. The signal $\bx(t)$ is thus of the form
$$\bx(t)=(x^1(t),{\cdots},x^M(t)),\qquad t\in\RR.$$
Each component $x^i(t)$ is then processed through an integrate-and-fire encoding machine, which outputs a sequence of spikes located at some increasing time instants $(t_j^i)_{j\in\Z_i}$, where $\Z_i$ is some index set of consecutive integers. From the derivations of \cite{Lazar04}, this provides the knowledge of successive integral values
\begin{equation}\label{sij}
\s_{i,j}:=\int_{t^i_{j-1}}^{t^i_j}x^i(t)\,\dif t,\qquad j\in\Z_i.
\end{equation}
The work of \cite{Adam20,Adam21} uses a POCS iteration to retrieve $\bx(t)$, before $\by(t)$ is recovered with the relation
\begin{equation}\label{yx}
\forall t\in\RR,\qquad\by(t)=\A\!^+\bx(t)
\end{equation}
where $\A\!^+$ is the matrix pseudo-inverse of $\A$.

\subsection{Signal and system formalization}\label{subsec:sys-formal}

We now show the existence of spaces $\scH$ and $\scA$ that allow to present the samples $\s_{i,j}$ of \eqref{sij} in the form of \eqref{sample0} with condition \eqref{cond0}.
Let $\scH:=(L^2(\RR))^M.$
Each element $\bu\in\scH$ is a function of time $\bu(t)=(u^1(t),{\cdots},u^M(t))$. The canonical inner product of $\scH$ is defined by
\begin{equation}\label{multi-inprod}
\langle\bu,\bv\rangle:={\textstyle\sum\limits_{i=1}^M}\langle u^i,v^i\rangle_2=\int_\RR\bu(t)^{\!\top}\bv(t)\,\dif t,\qquad\bu,\bv\in\scH
\end{equation}
where in the last expression, $\bu(t)$ and $\bv(t)$ are seen for each $t\in\RR$ as $M$-dimensional column vectors.
The signal $\bx(t)$ to be retrieved is an element $\bx\in\scH$ that specifically lies in the closed subspace
$$\scA:=\big\{\bv\in\scB^M:\forall t\in\RR,~\bv(t)\in\ran(\A)\big\}.$$
Next, let us show that $\s_{i,j}$ of \eqref{sij} can be formalized as
\begin{equation}\label{sij-formal}
\s_{i,j}=\big\langle\bx,\bg_{i,j}\big\rangle,\qquad(i,j)\in\Z.
\end{equation}
Naturally,
$$\Z:=\big\{(i,j):i\in\M \mbox{ and } j\in\Z_i\big\}~~\mbox{where}~~
\M:=\{1,\cdots,M\}.$$
Then, \eqref{sij-formal} clearly coincides with \eqref{sij} by taking
\begin{equation}\label{gij}
\bg_{i,j}(t):=\big(0,{\cdots},0,g_j^i(t),0,{\cdots},0\big)
\end{equation}
where $g_j^i(t)$ is at the $i$th coordinate position and is equal to
\begin{equation}\label{g-ASDM}
g_j^i(t)=1_{I_j^i}(t)\qquad\mbox{where}\qquad I^j_i:=[t_{j-1}^i,t_j^i].
\end{equation}
It is clear as a result that $(\bg_{i,j})_{(i,j)\in\Z}$ is an orthogonal family of $\scH$. Thus, condition \eqref{cond0} is realized. In fact, to obtain this property, it is sufficient to have
\begin{equation}\label{gij-ortho}
(g_j^i)_{j\in\Z_i} \mbox{  orthogonal in } L^2(\RR)\mbox{ for each } i\in\M.
\end{equation}
We will just assume this condition for now. and will apply the explicit assumption of \eqref{g-ASDM} only later on in Section \ref{subsec:g-ASDM}.

\subsection{POCS iteration implementation}

All of Sections \ref{sec:ortho-POCS} and \ref{sec:POCSlim} is applicable. In this process, we are thus uncovering the general pseudo-inversion properties of the POCS iteration limit of \eqref{x-iter3}, which was only studied in a noise-free case of perfect reconstruction with $\lambda_n=1$ in \cite{Adam20b,Adam21}. But the more specific contribution of interest here is the application of Section \ref{subsec:discrete} on a practical implementation of the POCS iteration.
We recall that the relaxed POCS algorithm of \eqref{x-iter-relax} is efficiently implemented by iterating the system \eqref{sys}.
Using the specific function notation of Section \ref{subsec:sys-formal}, this system is
\begin{subequations}\label{sys2}
\begin{align}
\vc\up{n+1}&=\vc\up{n}-\lambda_n S S^*\vc\up{n}+\vs_0\label{discr-iter2}\\
\bu\up{n}&=\bu\up{0}+S^*\vc\up{n}\label{xc2}
\end{align}
\end{subequations}
starting from $\vc\up{0}=\bzero$, where
\begin{subequations}\label{param}
\begin{align}
\vs_0&:=\lambda_n(\vs-S\bu\up{0}),\label{param-a}\\
 S S^*&=\begin{bmatrix}\displaystyle\frac{\langle {\widetilde\bg}_{i'j'},\bg_{i,j}\rangle}{\|\bg_{i'j'}\|^2}\end{bmatrix}_{((i,j),(i',j'))\in\Z\times\Z},\label{param-b}\\[0.5ex]
 S^*\vc&=\ssum{(i,j)\in \Z}\midfrac{\s_{i,j}}{\|\bg_{ij}\|^2}\,\widetilde{\bg}_{i,j},\quad\vc=(\s_{i,j})_{(i,j)\in\Z}\in\scD\label{param-c}\\
 \hspace{-3mm}\mbox{and}~~\widetilde{\bg}_{i,j}&=P_\scA\,\bg_{i,j},\quad(i,j)\in\Z.\label{param-d}
\end{align}
\end{subequations}
In the next two subsections, we derive the expressions of $SS^*$ and $S^*\vc$ n terms of the family of scalar functions $(g^j_i)_{(i,j)\in\Z}$.

\subsection{Discrete-time iteration}

The derivation of  $SS^*$ needed in \eqref{discr-iter2} starts with the following preliminary result.
\line
\begin{proposition}\label{prop:PA-mult}
If $\bu(t)=u(t)\,\vu$ where $u(t)\in L^2(\RR)$ and $\vu\in\RR^M$, then,
\begin{equation}\nonumber
P_\scA\bu(t)=\tilde u(t)\,\A\A\!^+\vu.
\end{equation}
\end{proposition}
\ppnoi
We prove this in Appendix \ref{app:PA-mult}. Then, the coefficients of the operator $SS^*$ described in \eqref{param-b} are given as follows.
\line
\begin{proposition}\label{prop:bg-inner}
For all $(i,j),(i',j')\in\Z$,
\begin{equation}\label{eq:bg-inner}
\|\bg_{i,j}\|^2={\|g_j^i\|}_2^2\quad\mbox{and}\quad\big\langle\widetilde{\bg}_{i',j'},\bg_{i,j}\big\rangle=
{\langle\tilde g^{i'}_{j'},g^i_j\rangle}_{\!2}\,\a_{ii'}
\end{equation}
where $\tilde u:=P_\scB\,u$ for any $u\in L^2(\RR)$, and $\a_{ii'}$ is the entry of matrix $\A\A\!^+$ at index $(i,i')\in\M{\times}\M$.
\end{proposition}
\line
\begin{IEEEproof}
The first equality of \eqref{eq:bg-inner} is clear from \eqref{gij} and \eqref{multi-inprod}. Let $\ve_i$ designate the $i$th coordinate vector of $\RR^M$. It follows from \eqref{gij}, \eqref{param-c} and Proposition \ref{prop:PA-mult} that
\begin{equation}\label{gtg}
\bg_{i,j}(t)=g_j^i(t)\,\ve_i\qquad\mbox{and}\qquad
\widetilde{\bg}_{i,j}(t)=\tilde g_j^i(t)\,\A\A\!^+\ve_i
\end{equation}
for all $(i,j)\in\Z$.
Thus, both $\bg_{i,j}(t)$ and $\widetilde{\bg}_{i,j}(t)$ are of the form $u(t)\,\vu$.
The general identity
$$\big\langle u(t)\vu,v(t)\vv\big\rangle=\langle u,v\rangle_2\,\vu^{\!\top}\vv$$
that results from \eqref{multi-inprod} then implies the second equality of \eqref{eq:bg-inner} with
$\a_{ii'}:=(\A\A\!^+\ve_i)^\top\ve_{i'}=\ve_i^\top(\A\A\!^+)\ve_{i'}$ since $\A\A\!^+$ is an orthogonal projection and hence symmetric.
\end{IEEEproof}

\subsection{Final continuous-time output}\label{subsec:output}

Once \eqref{discr-iter2} has been iterated the desired number of times $m$, one can output the continuous-time multi-channel signal $\bu\up{m}(t)$ from \eqref{xc2}. For that purpose, we need to know the explicit expression of $S^*\vc$ in terms of $\vc=(\s_{i,j})_{(i,j)\in\Z}\in\scD$. It follows from \eqref{param-c}, \eqref{eq:bg-inner} and \eqref{gtg} that $S^*\vc$ is the continuous-time function
\begin{eqnarray}
&\hspace{0mm}S^*\vc=\ssum{(i,j)\in \Z}\midfrac{\s_{i,j}}{\|g_j^i\|_2^2}\,\tilde g_j^i(t)\A\A\!^+\ve_i=\ssum{i\in \M}P_\scB(c^i(t))\A\A\!^+\ve_i\nonumber\\[0.5ex]
\lefteqn{\mbox{where}}&
c^i(t):=\ssum{j\in\Z_i}\midfrac{\s_{i,j}}{\|g_j^i\|_2^2}\,g_j^i(t).\label{ci}
\end{eqnarray}
If one needs to provide an estimate of the source signal $\by(t)$, the relation \eqref{yx} naturally leads us to consider the estimate
$$\bv\up{m}(t):=\A\!^+ \bu\up{m}(t)=\A\!^+ S^*\vc\up{m}\in\scB^N.$$
For any $\vc\in\scD$, it results from \eqref{ci} that
$$\A\!^+ S^*\vc=\ssum{i\in \M}P_\scB(c^i(t))\,\va^+_i$$
where $\va^+_i:=\A\!^+\A\A\!^+\ve_i=\A\!^+\ve_i$ as a general result of matrix pseudo-inverse. Then, $\va^+_i$ is nothing but the $i$th column vector of $\A\!^+$.

\subsection{Explicit case of samples of \eqref{sij}}\label{subsec:g-ASDM}

We now look at the more specific implementation for the samples $\s_{ij}$ considered in \eqref{sij}. As shown in  Section \ref{subsec:sys-formal}, this corresponds to the case where $g^i_j(t)$ is given by \eqref{g-ASDM}. The quantities that need to be derived more explicitly are
$${\|g_j^i\|}_2,\qquad{\langle\tilde g^{i'}_{j'},g^i_j\rangle}_{\!2}\qquad\mbox{and}\qquad c^i(t)$$
for \eqref{eq:bg-inner} and \eqref{ci}. It is first clear that
$$\|g_j^i\|_2^2=t_j^i-t_{j-1}^i.$$
As a generalization of a derivation from \cite{Thao21a} in the case of a single channel, it can be derived that
\begin{equation}\label{eq:inprod}
{\big\langle\tilde g^{i'}_{j'},g^i_j\big\rangle}_{\!2}=
f( T^{i,i'}_{j,{j'-1}})-f( T^{i,i'}_{j-1,{j'-1}})-f( T^{i,i'}_{j,j'})+f( T^{i,i'}_{j-1,j'})
\end{equation}
where $T^{i,i'}_{j,j'}:=t^i_j-t^{i'}_{j'}$ and
$f(t):=\int_0^t(t{-}\tau)\, \sinc(\tau)\,\dif\tau$.
Although ${\langle\tilde g^{i'}_{j'},g^i_j\rangle}_{\!2}$ depends on the four time instants $(t^i_j,t^i_{j-1},t^{i'}_{j'},t^{i'}_{j'-1})$ and is composed of four terms, it is the same single-argument numerical function $f(t)$ that is used. The values of this function can be stored in a lookup table, so that the required computation is just limited to a few additions and subtractions.

Finally
$$c^i(t)=\ssum{j\in\Z_i}\s_j^i\;1_{I_j^i}(t)\qquad\mbox{where}\qquad
\s_j^i:=\midfrac{\s_{i,j}}{t_j^i-t_{j-1}^i}.$$
This is nothing but the piecewise constant function equal to $\s_j^i$ in $I_j^i$ for each $j\in\Z_i$. This is produced by analog circuits using a zero-order hold.

\section{Bandlimited interpolation by iterative piecewise-linear corrections}\label{sec:Groch}

Until now, the examples of orthogonal sampling kernels we have provided all result from sampling by integration. An ultimate goal would be to have such a type of kernels for point sampling. For $x(t)$ that is at least continuous on $\RR$, one wishes to have $x(t_k)=\langle x,h_k\rangle_2$ where the functions $(h_k)\inZ$ have non-overlapping supports so that they are orthogonal in $L^2(\RR)$. The only way to do that would be to take $h_k(t):=\delta(t{-}t_k)$ where $\delta(t)$ is the Dirac impulse. However, this is not a function of $L^2(\RR)$. We show in this section that orthogonal kernels can actually be obtained for point sampling by choosing as ambient space $\scH$ a homogeneous Sobolev space. The resulting POCS method  turns out to coincide with an existing algorithm by Grochenig \cite[\S4.1]{Grochenig92b}. However, as in the example of Section \ref{sec:multi-TEM}, this algorithm was only analyzed in a case of noise-free perfect reconstruction, while our analysis leads to a complete set of pseudo-inversion properties.

\subsection{Initial idea}\label{subsec:idea}

Assume that  a bandlimited function $x$ is given by point samples $(x(t_k))\inZ$ at some known increasing sequence of instants $(t_k)\inZ$. For functions $u(t)$ of sufficient regularity, the point samples $(u(t_k))\inZ$ yield the relation
\begin{equation}
u(t_k){-}u(t_{k-1})=\int_{t_{k-1}}^{t_k}\!\!u'(t)\,\dif t=\left\langle u',1_{[t_{k-1},t_k]}\right\rangle_2=\left\langle u',g'_k\right\rangle_2\label{d-inner0}
\end{equation}
where
\begin{equation}\label{gk-ramp}
g_k(t):=\mbox{\small$\left\{\begin{array}{ccl}
0&,& t\in(-\infty,t_{k-1})\\[0.5ex]
t-t_{k-1} & , & t\in[t_{k-1},t_k)\\
t_k\!-t_{k-1}&,& t\in[t_k,\infty)
\end{array}\right.$},\quad k\in\Z.
\end{equation}
Using the notation
\begin{equation}\label{Sob-inner}
\Delta a_k:=a_k\!-a_{k-1}\qquad\mbox{and}\qquad\langle u,v\rangle:=\langle u',v'\rangle_2,
\end{equation}
we obtain from \eqref{d-inner0} that
\begin{equation}\label{d-inner}
\Delta u(t_k)=\langle u,g_k\rangle,\qquad k\in\Z.
\end{equation}
From the point samples $(x(t_k))\inZ$, we can then form  the  generalized samples
\begin{equation}\label{d-samp}
\s_k=\langle x,g_k\rangle\qquad\mbox{where}\qquad\s_k:=\Delta x(t_k),\qquad k\in\Z.
\end{equation}
Moreover, since $g'_k=1_{[t_{k-1},t_k]}$,  $(g'_k)\inZ$ is an orthogonal family in $L^2(\RR)$. So $(g_k)\inZ$ is orthogonal with respect to $\langle\cdot,\cdot\rangle$. The main issue is to find a Hilbert space $\scH$ in which  $\langle\cdot,\cdot\rangle$ is a well defined inner product.

\subsection{Ambient Hilbert space}

A suitable candidate for $\scH$ is the homogeneous Sobolev space \cite{grafakos2014modern,devore1993constructive} defined by
\begin{align*}
\sob=\Big\{u(t):u \mbox{ is absolutely continuous on } \RR\\[-1.5ex]
\mbox{and } u'\in L^2(\RR)&\Big\}
\end{align*}
where absolute continuity is here in the local sense. In this case, $u$ is absolutely continuous on $\RR$ if and only if it is differentiable almost everywhere of locally integrable derivative $u'$ such that
$u(b)=u(a)+\int_a^b u'(t)\,\dif t$ for any $a\leq b$ \cite[\S11.4.6]{berberian2012first}.
Since $u'\in L^2(\RR)$ for any $u\in\sob$, then the function $\langle\cdot,\cdot\rangle$ of \eqref{Sob-inner} is well defined in $\sob$. The remaining issue is that the induced function
\begin{equation}\label{Sob-norm}
\|u\|:=\langle u,u\rangle^{1\!/2}=\|u'\|_2
\end{equation}
is only a seminorm, as $\|u\|=0$ only implies that $u(t)$ is a constant function. In the construction of $\sob$, it is in fact implied that its functions are uniquely defined up to a constant component (similarly to the functions of $L^2(\RR)$ that are uniquely defined pointwise only up to a set of measure 0). Under this setting, $\|\cdot\|$ is a norm and $\big(\sob,\langle\cdot,\cdot\rangle\big)$ is rigorously a Hilbert space. Qualitatively, $\|u\|^2$ is the total slope energy of $u(t)$. We have the following list of properties.
\line
\begin{proposition}
\begin{enumerate}[label=(\roman*)]
\item $(g_k)\inZ$ is an orthogonal family of $\sob$.
\item For all $u\in\sob$, $u$ satisfies \eqref{d-inner}.
\item The subspace $\scA$ of all bandlimited functions of $\sob$ of Nyquist period 1 is closed.
\end{enumerate}
\end{proposition}
\line
\begin{IEEEproof}
(i) As we already saw the orthogonality of $(g_k)\inZ$ with respect to $\langle\cdot,\cdot\rangle$, we just need to verify that $g_k\in\sob$. For each $k$, $g_k$ is easily seen to be  absolutely continuous. Meanwhile,  $\|g'_k\|_2^2=\|1_{[t_{k-1},t_k]}\|_2^2=\Delta t_k$ so that $g'_k\in L^2(\RR)$.

(ii) By absolute continuity, every function $u\in\sob$ satisfies \eqref{d-inner0} and hence \eqref{d-inner}.

(iii) Let $U(\omega)$ be the Fourier transform of a function $u(t)$. For any interval $I$, let $u_I(t)$ be the function whose Fourier transform is $1_I(\om)U(\om)$. For any given $u\in\sob$, we can write $u=u_B+u_{\bar B}$ where $B:=[-\pi.\pi]$ and $\bar B:=\RR\backslash B$. Since $\|v'\|_2^2=\frac{1}{2\pi}\int_\RR\omega^2\,|V(\omega)|^2\,\dif\omega$, it is clear that $u_B',u_{\bar B}'\in L^2(\RR)$.  Meanwhile, they are both  absolutely continuous since $u_B$ is infinitely differentiable and $u_{\bar B}=u-u_B$. So $u_B,u_{\bar B}\in\sob$. Since
$\langle v,w\rangle=\frac{1}{2\pi}\int_\RR\omega^2\, V(\omega)W^*(\omega)\,\dif\omega$,
then $u_B\perp u_{\bar B}$ in $\sob$. While $\scA$ is the subset of $\sob$ of functions whose Fourier transforms are supported by $B:=[-\pi,\pi]$, let $\bar\scA$ be  the subset of $\sob$ of functions whose Fourier transforms are supported by $\bar B$. We have just proved that $\scA\oplus\bar\scA$ is an orthogonal decomposition of $\sob$. This proves that $\scA$ is closed in $\sob$.
\end{IEEEproof}

\subsection{POCS algorithm}\label{subsec:sob-POCS}

After forming the generalized samples $\s_k:=\Delta x(t_k)$ from the point samples of $x$, we have shown in \eqref{d-samp} that the generic sampling form of \eqref{sample0} is achieved with $\scH=\sob$, $\scA=\scB$ and the family $(g_k)\inZ$ defined in \eqref{gk-ramp} which is orthogonal in $\scH$.
All the conditions of Section \ref{sec:ortho-POCS} are thus realized. The  signal $x$ can then be estimated by iterating
\begin{equation}\label{x-iter2}
u\up{n+1}=P_\scA P_{\scH_\vs} u\up{n},\qquad n\geq0
\end{equation}
which we have simply repeated from \eqref{x-iter1} for convenience, where $\scH_\vs$ is defined in \eqref{Hs} and $\vs:=(\Delta x(t_k))\inZ$.
\line
\begin{proposition}\label{prop:Sob-Hs}
With $\vs:=(\Delta x(t_k))\inZ$, $\scH_\vs$ is the set of functions $u(t)\in\sob$ such that $u(t_k)-x(t_k)$ is constant for $k\in\Z$.
\end{proposition}
\line
\begin{IEEEproof}
From \eqref{d-inner}, $\langle u,g_k\rangle=\Delta u(t_k)$ for any $u\in\sob$. So, from \eqref{Hs},
$u\in\scH_\vs$ if and only if $\Delta u(t_k)=\s_k=\Delta x(t_k)$ for all $k\in\Z$. This is equivalent to $u(t_k)-x(t_k)=u(t_{k-1})-x(t_{k-1})$ for all $k\in\Z$. This proves the proposition.
\end{IEEEproof}
\ppnoi
By applying the POCS results of Section \ref{subsec:ortho-POCS} to \eqref{x-iter2}, we obtain the following result.
 \line
\begin{proposition}\label{prop:Groch-conv}
Assume that $(t_k)\inZ$ is any increasing sequence of time instants. Let $(u\up{n})_{n\geq0}$ be recursively defined by \eqref{x-iter2} starting from some $u\up{0}\in\scA$, and $u\up{\infty}$ be the function of $\scA$ that interpolates the points $(t_k,x(t_k))\inZ$ while minimizing $\|(u\up{\infty}\!-u\up{0})'\|_2$. Then, $\|(u\up{n}\!-u\up{\infty})'\|_2$ monotonically tends to 0 with $n$.
\end{proposition}
\line
\begin{IEEEproof}
 Given that $\scH=\sob$ with the norm defined in \eqref{Sob-norm}, we know from Section \ref{subsec:ortho-POCS} that $\|u\up{n}\!-\hat u\|=\|(u\up{n}\!-\hat u)'\|_2$ monotonically tends to 0 with $\hat u:=P_{\scA_\vs}u\up{0}$. Since $\hat u\in\scA_\vs=\scA\cap\scH_\vs$, we know by Proposition \ref{prop:Sob-Hs} that $\hat u$ is an element of $\scA$ that interpolates the points $(t_k,x(t_k))\inZ$ up to a constant component. But since $\hat u=P_{\scA_\vs}u\up{0}$, it also minimizes $\|\hat u-\hat u\up{0}\|=\|(\hat u-\hat u\up{0})'\|_2$. Thus $\hat u$ and $u\up{\infty}$ differ by just a constant. Then, $\hat u$ can be replaced by $u\up{\infty}$  in all the above norm expressions.
\end{IEEEproof}

\subsection{Conincidence with Grochenig's algorithm}

As $\vs=(\Delta x(t_k))\inZ$ is equivalent to $x\in\scH_\s$, we recall from Proposition \ref{prop:proj} that
\begin{equation}\label{PGs2}
P_{\scH_\vs}u:=u+P_\scG(x{-}u)
\end{equation}
where $\scG$ is the closed linear span of $(g_k)\inZ$. Given the definition of this family in \eqref{gk-ramp}, $P_\scG$ is characterized as follows.
\line
\begin{proposition}\label{prop:PL}
For any $u\in\sob$, $P_\scG u$ is the function that linearly interpolates the points $(t_k,u(t_k))\inZ$ up to a constant component.
\end{proposition}
\line
\begin{IEEEproof}
Let $u\in\sob$. Since $(g_k)\inZ$ is an orthogonal basis of $\scG$, then
\begin{equation}\label{PG-lin}
P_\scG u=\smallsum{k\in\Z}\midfrac{\langle u,g_k\rangle}{\|g_k\|^2}\,g_k(t)
=\smallsum{k\in\Z}\midfrac{\Delta u(t_k)}{\Delta t_k}\,g_k(t)
\end{equation}
using \eqref{d-inner} and the result $\|g_k\|^2=\|1_{[t_{k-1},t_k]}\|_2^2=\Delta t_k$ .
Let $\hat u$ be the linear interpolant of the points $(t_k,u(t_k))\inZ$. For each $k\in\Z$ and every $t\in(t_{k-1},t_k)$, it is easy to see that $u'(t)=\Delta u(t_k)/\Delta t_k=\hat u'(t)$. As $u$ and $\hat u$ are both absolutely continuous, then $u-\hat u$ is a constant function.
\end{IEEEproof}
\ppnoi
Grochenig previously introduced in \cite[\S4.1]{Grochenig92b} the following iteration\footnote
{This iteration appears in eq.(24) of \cite{Grochenig92b} in the equivalent form of $u\up{n+1}=u\up{n}+P_\scB P_\scG(x{-}u\up{n})$ since $u\up{n}\in\scB$.}
\begin{equation}\label{Groch-iter}
u\up{n+1}=P_\scB\big(u\up{n}\!+L(x{-}u\up{n})\big),\qquad n\geq0
\end{equation}
where $u\up{n}$ is in the subspace $\scB\subset L^2(\RR)$ of bandlimited functions, which we assume of Nyquist period 1, and $Lu$ is the exact linear interpolation of the points $(t_k,u(t_k))\inZ$. When thinking of $u\up{n}$ as elements of $\scA\subset\sob$, then \eqref{Groch-iter} coincides with \eqref{x-iter2} given \eqref{PGs2}.

\subsection{Analysis comparison}

It is interesting to compare the convergence analysis of \eqref{Groch-iter} from \cite{Grochenig92b} with the result of Proposition \ref{prop:Groch-conv}.
Under the condition that
\begin{equation}\label{Groch-cond}
\lim_{k\rightarrow\pm\infty}t_k=\pm\infty\qquad\mbox{and}\qquad
\Delta<1,
\end{equation}
where $\Delta:=\textstyle\sup_{k\in\ZZ}\,\Delta t_k$,
it was shown in \cite{Grochenig92b} that the transformation of \eqref{Groch-iter} is a contraction with respect to the $L^2$-norm. As $x$ is a fixed point of \eqref{Groch-iter}, this proves that $u\up{n}$ linearly converges in $L^2$-norm to $x$. Meanwhile, Proposition \ref{prop:Groch-conv} analyzes the convergence of $u\up{n}$ in terms of the $L^2$-norm of its derivative. A shortcoming is the loss of information on its constant component. However, this proposition contains a lot more results, as follows.

Note first that none of the conditions of \eqref{Groch-cond} are assumed in Proposition \ref{prop:Groch-conv}. In the case where $t_k$ has a finite limit $t_\infty$ (resp. $t_{-\infty}$) when $t$ goes to $\infty$ (resp. $-\infty$), then we just need to assume that $Lu$ is constant and equal to $u(t_\infty)$ (resp. $u(t_{-\infty})$) in $[t_\infty,\infty)$  (resp. $(-\infty,t_{\infty}$]) to be consistent with $P_\scG u$ as a result of \eqref{PG-lin} with \eqref{gk-ramp}. We can even include the case where $\Z$ is a finite set $\{1,{\cdots},N\}$, which corresponds to the case of $N{+}1$ sampling instants $t_0,\cdots,t_N$ ($t_0$ and $t_N$ playing the roles of $t_{-\infty}$ and $t_\infty$ in the construction of $L$).

 Next, we can see that perfect reconstruction is achieved (up to a constant component) whenever the samples uniquely define $x(t)$ as a bandlimited signal. There are known cases where this is realized without the constraint that $\Delta<1$. For example, uniqueness of reconstruction is known from \cite{Duffin52} to be realized when $(|t_k{-}kT|)_{k\in\ZZ}$ is upper bounded for some $T\in[0,1)$ and $\Delta t_k$ has a positive lower bound. This condition can lead to arbitrarily large $\Delta$.

 When consistent reconstruction is not unique, Proposition \ref{prop:Groch-conv} implies that \eqref{Groch-iter} is still convergent (up to a constant component) and the limit $u\up{\infty}$ is the bandlimited interpolator that is closest to $u\up{0}$ in terms of the norm of \eqref{Sob-norm}. If $u\up{0}=0$, $u\up{\infty}$ can be interpreted as the bandlimited interpolator of minimum slope energy. But choosing a nonzero function $u\up{0}$ can be useful to ``attract'' the interpolator $u\up{\infty}$ towards some signal guess obtained by other means (see experiment in Section \ref{subsec:exp2}).

On top the uncertainty on constant component, a shortcoming of Proposition \ref{prop:Groch-conv} is the absence of linear convergence. This would require to know the condition on $(t_k)\inZ$ for $\ran(S)$ to be closed. This is a difficult problem that cannot be resolved here. However, we know by default that linear convergence is achieved whenever  $\Z$ is finite, which is the case of interest in practice. Finally, our framework includes the additional option of relaxation and insight on the POCS limit under sampling noise.

The lack of knowledge on the constant components is mostly due to the analysis in $\sob$ which cannot incorporate the fact that $L$ performs in \eqref{Groch-iter} an exact interpolation. It is actually possible to combine the consequences of this analysis with the latter fact and eliminate the constant-component uncertainties. As this is beyond the main scope of the present paper, this will be reserved for a future publication.

\subsection{Numerical experiments in oversampling situation}\label{subsec:exp1}

\begin{figure*}
\begin{centering}
\hbox{
\vbox{\hbox{\includegraphics[width=0.67\columnwidth]{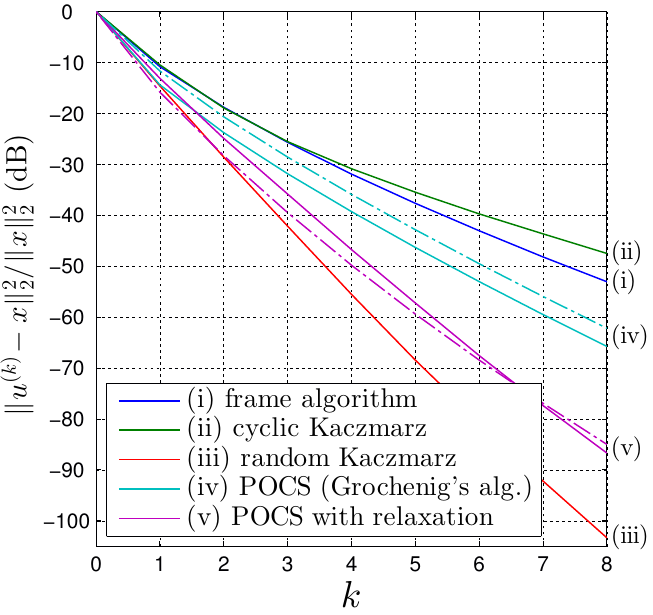}}
\hbox{\makebox[2.35in]{(a)}}}
\hspace{0.mm}\vbox{\hbox{\includegraphics[width=0.67\columnwidth]{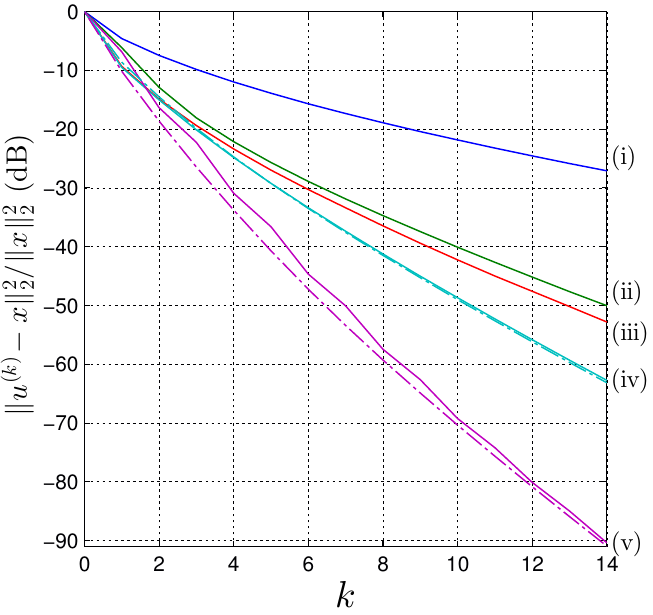}}
\hbox{\makebox[2.35in]{(b)}}}
\hspace{0.mm}\vbox{\hbox{\includegraphics[width=0.67\columnwidth]{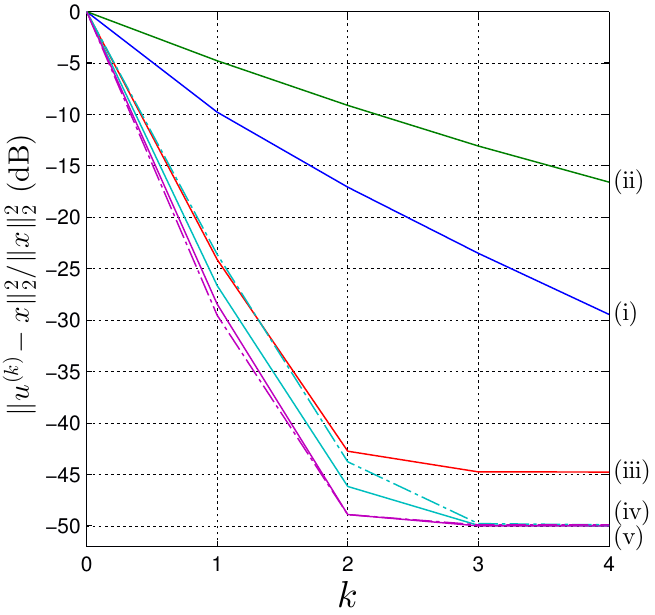}}
\hbox{\makebox[2.35in]{(c)}}}}
\end{centering}
\vspace{0mm}
\caption{MSE results of iterative bandlimited reconstructions from nonuniform point samples at randomly generated instants $(t_k)\inZ$ with  the following statistics: (a) uniform distribution of $\Delta t_k$ in $[0.3,1]$ (1.3 relaxation for (v)); (b) randomly positioned clusters of sampling instants (1.45 relaxation for (v)); (c) uniform distribution of $\Delta t_k$ in $[0,0.5]$ with sampling noise (1.05 relaxation for (v)). The solid lines correspond to MSE based on the $L^2$-norm (as reported in the ordinate description), while the mixed lines correspond to MSE based the Sobolev norm.}\label{fig2}
\vspace{0mm}
\end{figure*}
We plot in Fig. \ref{fig2} the MSE performance of a number of iterative bandlimited reconstruction algorithms from nonuniform point samples of similar complexity per iteration, including the original frame algorithm introduced in the pioneering paper on nonuniform sampling \cite{Duffin52}, the basic Kaczmarz method and its randomized version presented in Section \ref{sub:Kaczmarz}, and Grochenig's algorithm without and with relaxation. For each algorithm, the MSE of the $n$th iterate $u\up{n}$, reported in solid lines, is measured by averaging the relative error $\|u\up{n}\!-x\|_2^2/\|x\|_2^2$ over 100 randomly generated bandlimited inputs $x$ that are periodic of period 315 (assuming a Nyquist period 1). Even though our analysis of Grochenig's algorithm has been constructed in the Sobolev space $\sob$, we have maintained the $L^2$-norm in the error measurements as this is the standard reference of MSE in signal processing. We have however superimposed in mixed lines the MSE obtained by averaging $\|u\up{n}\!-x\|^2/\|x\|^2$ where $\|\cdot\|$ is the norm of $\sob$ given in \eqref{Sob-norm}, specifically for the results of Grochenig's algorithm in curves (iv) and (v). Even though the two norms are not equivalent, we observe that they yield similar results. So, while we do not provide analytical justifications for this similarity, we see that the Sobolev norm remains an adequate tool of error predictions in these experiments. In (v), the relaxation coefficient has been optimized empirically.

In Fig. \ref{fig2}\,(a,b), we compare the behavior of the algorithms with respect to two types of sampling nonuniformity. In (a), $(\Delta t_k)\inZ$ is generated as an i.i.d sequence that is uniformly distributed in $[0.3,1]$ (in Nyquist period unit), leading to an oversampling ratio of about 1.54. Meanwhile, the sampling instants in (b) are grouped into clusters that are nonuniformly spaced between each other. Each cluster is made of 3 sampling instants equally spaced by $1/4$ (as opposed to 1 in Nyquist rate sampling), and the overall density of of clusters is such that the oversampling ratio is $2$. The two figures show that the randomized Kaczmarz method appears to be well tuned for the random nonuniformity of (a), but not at all for the clustered type of nonuniformity in (b), where it barely improves the basic cyclic version of the Kaczmarz method. Meanwhile, Grochenig's algorithm shows its systematic superiority to the cyclic Kaczmarz method in MSE. This is particularly remarkable for the clustered type of nonuniformity, which is known to be a challenge for bandlimited interpolation. Now, the inclusion of relaxation in (v) is also of  outstanding impact: it substantially improves the unrelaxed version in a way that could not have been predicted or justified in the the general framework of contractions, like in the analysis of \cite{Grochenig92b}.

Fig. \ref{fig2}\,(c) goes back to a nonuniformity of the type of (a) at however a higher oversampling ratio to highlight the noise-shaping effect of the POCS algorithm in the presence of sampling errors. Here, the sequence $(\Delta t_k)\inZ$ is uniformly distributed in $[0,0.5]$, and the sample errors are Gaussian random variables that are 45 dB's below the input in variance. While the randomized Kaczmarz method exhibits the same type of fast convergence as in (a), it shows inferior capabilities of noise filtering compared to Grochenig's method. This is a consequence of the pseudo-inversion property of the latter method when formalized as a POCS algorithm. As an extra result, the figure shows the particularly poor behavior of the cyclic Kaczmarz method to high oversampling.

\subsection{Numerical experiments in sub-Nyquist situation}\label{subsec:exp2}

\begin{figure}
\centerline{\hbox{\scalebox{0.7}{\includegraphics{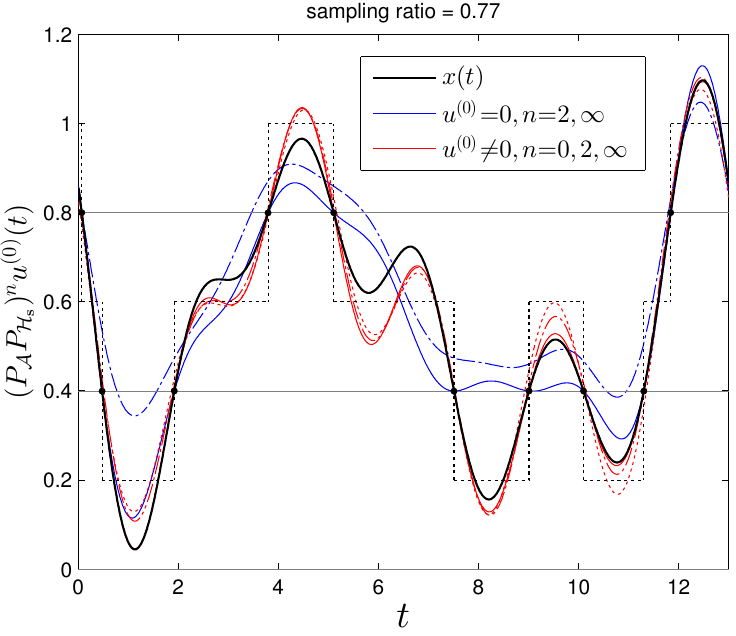}}}}
\vspace{-1mm}
\caption{Estimation of bandlimited input $x(t)$ from level crossing samples in sub-Nyquist situation (sampling ratio of $0.77$).}\label{fig3}
\vspace{0mm}
\end{figure}
We show in Fig. \ref{fig3} an example of POCS iteration limit in a case of sub-Nyquist sampling. In this experiment, the samples of the bandlimited input $x(t)$ are obtained by level-crossing sampling [6], [8], [9], that is, from the crossings of $x(t)$ with fixed levels represented by horizontal grey lines \cite{Marvasti01,Miskowicz2018,Rzepka18}. The resulting sampling ratio is $0.77$ (the time unit is the Nyquist period of $x(t)$), which prevents uniqueness of reconstruction. While the blue curves result from a zero initial estimate $u\up{0}(t)$, the red curves are obtained by choosing for $u\up{0}(t)$ the bandlimited version of the piecewise constant function shown in black dotted curve. This stair case function can be generated from the mere knowledge of the level crossings. In each of the two cases of initial estimate, the result of 3 iteration numbers is plotted, using the following line types in increasing order of iteration: dotted line, mixed line, solid line. The exact iteration numbers of the plots are indicated in the legend. To validate our convergence analysis, we have actually plotted the result of infinite iteration by employing the theoretical limit formula of \eqref{x-inf3} and computing $S^\dagger$ by matrix pseudo-inverse, given the low dimensionality of the experiment. The figure shows the good convergence of $u\up{n}(t)$ to this ideal limit. Now, the main point of this figure is to show an example of reconstruction  improvement using a heuristically designed nonzero initial estimate $u\up{0}(t)$.

\section{Summary}

We have introduced in this paper an abstract framework where a bandlimited input signal can be  reconstructed/estimated from nonuniform generalized samples by pseudo-inversion of the sampling operation, using an algorithm that consists of iterated time-varying filters. This requires that the sampling kernel functions be orthogonal at least in a Hilbert space that is larger than the input space. We prove in this paper the full pseudo-inversion properties of our algorithm. They include perfect reconstruction whenever the samples uniquely define the input, minimum-norm reconstruction when the sampling is insufficient, and a noise-shaping effect on sampling errors. While the required condition on the sampling kernels was previously observed in the time encoder of Lazar and T\'oth, this condition is shown in this paper to appear in a non-trivial manner in a recent multi-channel time-encoding system as well as in traditional point sampling, Our resulting algorithms turn out to coincide with existing reconstruction methods in these cases, But our framework reveals the pseudo-inversion properties of these methods, while proposing efficient discrete-time implementations.

\appendix

\subsection{Proof of Theorem \ref{theo:relax}}\label{app:relax}

For any  closed affine subspace $\scC $ and $\lambda\in\RR$, $P_\scC ^\lambda u-u=\lambda(P_\scC u-u)\in{\overrightarrow\scC}^\perp$ where $\overrightarrow\scC$ is the linear subspace associated with $\scC$. Since $P_{\scH_\vs}^2 u-u=2(P_{\scH_\vs}u-u)$, $P_{\scH_\vs}^2$ is the symmetry with respect to the affine space $\scH_\vs$, and hence is non-expansive. Using the relation, $P_\scC ^\lambda u-u=\frac{\lambda}{2}(P_\scC ^2u-u)$, we have $P_\scA P_\scC ^\lambda u-u=\frac{\lambda}{2}(P_\scA P_\scC ^2u-u)$ for all $u\in\scA$.  By applying this with $\scC ={\scH_\vs}$, $\lambda=\lambda_n$ and $u=u\up{n}$, the iterate $u\up{n}$ of \eqref{x-iter3} satisfies the recursion
\begin{equation}\nonumber
u\up{n+1}=u\up{n}+\smallfrac{\lambda_n}{2}(R u\up{n}\!-u\up{n}),\qquad n\geq0
\end{equation}
where $R:=P_\scA P_{\scH_\vs}^2$. This operator is non-expansive and the set of its fixed points can be verified to be $\scA\cap\scH_\vs=\scA_\vs$. As $\sum_{n\geq0}\smallfrac{\lambda_n}{2}(1{-}\smallfrac{\lambda_n}{2})=+\infty$ with $\frac{\lambda_n}{2}\in[0,1]$, we can apply Theorems 5.14 and 5.13 of \cite{bauschke2017convex}, and conclude that $u\up{n}$ has a limit $u\up{\infty}\in\scA_\vs$.
We have $P_{\scH_\vs}^{\lambda_n}u\up{n}-u\up{n}\in{\overrightarrow\scH_\vs\!\!}^\perp$ while $P_\scA P_{\scH_\vs}^{\lambda_n}u\up{n}-P_{\scH_\vs}^{\lambda_n}u\up{n}\in\scA^\perp$. Therefore, $u\up{n+1}-u\up{n}\in\scA\!^\perp{+}{\overrightarrow\scH_\vs\!\!}^\perp\subset
{(\scA\cap\overrightarrow\scH_\vs)\!\!}^\perp={\overrightarrow\scA_\vs\!\!}^\perp$. Hence,
$u\up{\infty}\!-u\up{0}\!\in{\overrightarrow\scA_\vs\!\!}^\perp$. This proves that $u\up{\infty}\!=P_{\scA_\vs}u\up{0}$.

\subsection{Proof of Proposition \ref{prop:equiv1}}\label{app:equiv1}

(i) $\Leftrightarrow$ (ii): Given \eqref{S}, it is known from \cite{Kato95} that $\ran(S)$ is closed in $\scD$ if and only if there exists a constant $\alpha>0$ such that $\|Su\|_\scD\geq\alpha\|u\|$ for all $u\in\null(S)^\perp$. This amounts to (ii) with the identity $\null(S)^\perp=\scF$ from \eqref{nullS}.

(ii) $\Leftrightarrow$ (iii): By definition, $(f_k)\inZ$ is a frame of $\scF$ if and only if there exist constants $0<A\leq B$ such that
\begin{equation}\label{frameseq}
\forall u\in\scF,\qquad A\|u\|^2\leq\smallsum{k\in\Z}|\langle u,f_k\rangle|^2\leq B\|u\|^2
\end{equation}
Now, it follows from \eqref{inner-proj} and \eqref{Su-norm} that
\begin{equation}\label{prelim0}
\forall u\in\scA,~~\smallsum{k\in\Z}|\langle u,f_k\rangle|^2=\smallsum{k\in\Z}\midfrac{|\langle u,\widetilde g_k\rangle|^2}{\|g_k\|^2}=\|Su\|_\scD^2.
\end{equation}
We see from \eqref{Su-norm} that the upper bound of \eqref{frameseq} is by default satisfied with $B:=1$. Meanwhile, the lower bound $A>0$ exists if and only if (ii) is satisfied.

\subsection{Proof of Proposition \ref{prop:PA-mult}}\label{app:PA-mult}

\begin{lemma}
Let $\bu(t):=u(t)\,\vu$ where $u(t)\in L^2(\RR)$ and $\vu\in\RR^M$. If $u(t)\in\scB^\perp$ or $\vu\in\ran(\A)^\perp$, then $\bu(t)\in\scA^\perp$.
\end{lemma}
\line
\begin{IEEEproof}
Let $\bv(t)\in\scA$. It follows from \eqref{multi-inprod} that
$$\langle\bu,\bv\rangle=\big\langle u(t)\vu,\bv(t)\big\rangle=\big\langle u(t),\vu^{\!\top}\bv(t)\big\rangle_2.$$
Clearly, $\vu^{\!\top}\bv(t)\in\scB$. So if $u(t)\in\scB^\perp$, then $\langle\bu,\bv\rangle=0$. Meanwhile, if $\vu\in\ran(\A)^\perp$, then $\vu^{\!\top}\bv(t)=0$ for each single $t\in\RR$. Then  $\langle\bu,\bv\rangle=0$ regardless of $u(t)$. This proves the lemma.
\end{IEEEproof}
\ppnoi
We now proceed with the proof of Proposition \ref{prop:PA-mult}.
It will be convenient to define $\P:=\A\A\!^+$. Let $\bw(t):=\tilde u(t)\,\P\vu$. Its $i$th component is $w^i(t)=\tilde u(t)\,\q^i\in\scB$, where $\q^i$ is the $i$th coordinate of $\P\vu$. So $\bw(t)\in\scB^M$. Meanwhile, $\P\vu\in\ran(\A)$, so $\bw(t)\in\ran(\A)$ for each $t\in\RR$. Then, $\bw(t)\in\scA$. Next, we can write
$$\bu(t)-\bw(t)=\big(u(t){-}\tilde u(t)\big)\vu+\tilde u(t)\,(\vu{-}\P\vu).$$
While $u(t){-}\tilde u(t)\in\scB^\perp$, $\vu-\P\vu\in\ran(\A)^\perp$ because $\P$ is known to be the orthogonal projection of $\RR^M$ onto $\ran(\A)$. So $\bu(t)-\bw(t)\in\scA^\perp$ according to the above lemma. Thus, $\bw(t)=P_\scA\bu(t)$.

\section*{Acknowledgment}

The authors would like to thank Eva Kopeck\'a and Patrick Combettes for their indications on the convergence of relaxed POCS, and Sinan G\"unt\"urk for his help on Sobolev spaces.

\bibliographystyle{ieeetr}

\bibliography{reference}{}

\end{document}